\documentclass[aps,prb,preprint,showpacs,superscriptaddress,floatfix]{revtex4}
\usepackage{graphics}
\usepackage{epsfig}
\usepackage{amsmath}
\usepackage{amssymb}
\usepackage{calc}
\newcommand{\beq}{\begin{equation}}
\newcommand{\eeq}{\end{equation}}
\newcommand{\bea}{\begin{eqnarray}}
\newcommand{\eea}{\end{eqnarray}}

\begin{document}
\title{ Strings, black holes, the tripartite entanglement of seven qubits and the Fano plane }
\author{P\'eter L\'evay}
\affiliation{Department of Theoretical Physics, Institute of Physics, Budapest University of Technology and Economics, H-1521 Budapest, Hungary}
\date{\today}
\begin{abstract}
Recently it has been observed that the group $E_7$ can be used to describe a special type of quantum entanglement of seven qubits partitioned
into seven tripartite systems.
Here we show that this curious type of entanglement is entirely encoded into the discrete geometry of the Fano plane.
We explicitly work out the details concerning a qubit interpretation of the $E_7$ generators as representatives of tripartite protocols acting on 
the $56$ dimensional representation space.
Using these results we extend further the recently studied analogy between quantum information theory and supersymmetric black holes in four-dimensional string theory. We point out that there is a  dual relationship between entangled subsystems containing three and four tripartite systems. This relationship is reflected in the structure of  
the expressions for the black hole entropy in the $N=4$ and $N=2$ truncations 
of the $E_{7(7)}$ symmetric area form of $N=8$ supergravity.
We conjecture that a similar picture based on other qubit systems might hold for black hole solutions in magic supergravities.
\end{abstract}
\pacs{
03.65.Ud, 04.70.Dy, 03.67.Mn, 02.40.-k}
\maketitle{}
\section{Introduction}
Recently striking multiple relations have been established between two seemingly unrelated
strains of knowledge: quantum information theory and the physics of four-dimensional stringy black holes.
The activity in this field has started after a paper of Duff\cite{Duff}
pointing out a mathematical coincidence between the form of the macroscopic 
entropy for the four-dimensional BPS STU black hole and the three-tangle\cite{Kundu}
of three-qubit systems expressed in terms of Cayley's hyperdeterminant.
This mathematical coincidence is based on the similar symmetry properties one encounters in these different physical situations. As far as classical supergravity is concerned in the STU model the group representing the symmetry in question is $SL(2,{\bf R})^{\otimes 3}$ or, taking into account quantum corrections and the quantized nature of electric and magnetic charges
$SL(2, {\bf Z})^{\otimes 3}$. In models of qubit systems on the other hand the symmetry group is the one of stochastic local operations and classical communication\cite{Vidal} (SLOCC)
which is $SL(2,{\bf C})^{\otimes 3}$.
Later by looking deeper into the structure of such models further mathematical similarities have been found\cite{Linde}.
In their paper Kallosh and Linde have extended the validity of the relationship established between the three-tangle and the BPS STU black hole entropy to non-BPS ones.
They have also related the well-known entanglement classes  of pure three-qubit entanglement\cite{Vidal} to different classes of stringy black holes in $N=2$ supergravity.
Of course the appearance of similar mathematical structures in two disparate subjects does not necessarily imply a deeper unity however, the realization that these relations do exist might turn out to be important for obtaining further insights in both fields.  
Moreover, these mathematical coincidences in both of our theories are related to the same basic physical concepts namely entropy, information and entanglement
hence their occurrence in different physical situations deserves further investigation.
Following this guideline we have shown\cite{Levay} that the extremization of the BPS mass with respect to the {\it complex} moduli of the STU model is connected to the problem of finding an optimal distillation protocol of a GHZ state from an initial one defined by the charges and the moduli.
Alternatively, finding the frozen values of the moduli via the supersymmetric attractor mechanism\cite{attractor} is related to the quantum information         theoretic  
scenario of using a special type of quantum algorithm  for the maximization of tripartite entanglement.
The entangled three-qubit states occurring in this protocol  
are complex ones 
containing the quantized charges ($4$ electric and $4$ magnetic) {\it and } the complex moduli. However, it turns out that they are local unitary equivalent to real quantum bits ("rebits"), in accordance with the symmetry group $SL(2, {\bf R})^{\otimes 3}$ of the STU model.

Can we generalize this nice picture for more general types of four-dimensional black hole solutions?
The first indication that this generalization might be done came from Kallosh and Linde\cite{Linde}.
They have emphasized the universal role of the Cartan-Cremmer-Julia $E_{7(7)}$ invariant is playing as the expression for the entropy of black holes and black rings in the more general context of $N=8$ supergravity/M theory.
By making use of the $SU(8)$ symmetry present in these models they have shown that the three-tangle shows up in this invariant too.
However, the symmetry group in the $N=8$ case is $E_{7(7)}$ and in the $N=2$ STU one it is $SL(2, {\bf R})^{\otimes 3}$ hence in spite of this result  it is not at all obvious that
three-qubit systems are also relevant here. 
Indeed, as was remarked by Pioline\cite{Boris} 
according to Freudenthal's construction which underlies the derivation of the quartic Cartan-Cremmer-Julia invariant based on the split octonions
the electric and magnetic charges are associated with a square (with elements from ${\bf R}$ in the diagonal and Jordan algebra elements in the off-diagonal)   rather than a cube. 
Hence the three-qubit interpretation of the STU model might be difficult to generalize.

However, in a recent paper of Duff and Ferrara\cite{Ferrara} it turned out that a three-qubit interpretation of the $E_7$ symmetric black hole entropy formula is quite natural.
The trick is to take instead of a single three-qubit system  {\it seven} copies of them.
The fundamental $56$ dimensional representation of $E_7$ can then be built from
a suitable direct sum of such tripartite subsystems.
The authors also presented the qubit form of the quartic Cartan-Cremmer-Julia 
invariant $J_4$ which they regarded as a measure
which should characterize this unusual type of tripartite entanglement of seven qubits.
Finally they indicated how the important special cases of the $N=4$ and $N=2$ (STU model) supersymmetric black hole entropy formulas are incorporated into this formalism as special cases obtained by truncation.

The aim of the present paper is to investigate further the entanglement         properties
of this seven qubit system associated with the group $E_7({\bf C})$, to set the ground for establishing further possible relationships between quantum information theory and the physics of stringy black holes.
We show that this curious type of entanglement is entirely encoded into the
 discrete geometry of the Fano plane, the smallest projective plane.
We point out that the entanglement associated to $E_7({\bf C})$
is just the one of an entangled lattice defined by the incidence graph of the Fano plane.
We explicitly work out the details concerning a qubit interpretation of the     $E_7({\bf C})$
infinitesimal operators as representatives of tripartite protocols acting on
the $56$ dimensional representation space.
 Using these results we extend further the recently studied analogy between     quantum
information theory and supersymmetric black holes in four-dimensional string
theory. We point out that in this entangled lattice there is a  dual relationship between unnormalized  entangled
states 
containing three and four qubits, or using projective duality between subsystems containing three and four tripartite systems. This relationship is reflected
in the structure of
 the expressions for the black hole entropy in the $N=4$ and $N=2$ truncations
 of the $E_{7(7)}$ symmetric area form of $N=8$ supergravity.
  
The organization of the paper is as follows.
In section II. we consider four qubit systems. Here we regard a particular four-qubit state as an {\it operator} that is in turn acting on entangled states forming new entangled ones. We can organize these four-qubit states taken together with their associated SLOCC transformations acting on them into an $so(4,4,{\bf C})$ algebra.
Next using triality and the four-qubit picture we define three different types  of representations of $so(4,4,{\bf C})$ to be used later.
In section III. we show how the entangled system as discussed by Duff and Ferrara\cite{Ferrara} is just the one encoded into the Fano plane with qubits attached to its vertices.
Here we show that the different types of tripartite subsystems formed from the seven qubits can be characterized by a ${\bf Z}_2^3$ charge, hence the representation space for the fundamental of $E_7({\bf C})$ is built from the direct sum of different superselection sectors. 
Here an interesting connection with error correcting codes is also pointed out.
Section IV. is devoted to the construction of the generators of the Lie-algebra $e_7({\bf C})$. 
In Section V. we sketch the construction of the explicit form of the fundamental $56$ of $E_7$ in terms of tripartite protocols for seven qubits.
Cartan's invariant as a measure of entanglement is introduced in Section VI.
We show how its structure can be understood in terms of the geometric data of   the dual Fano plane.
In two separate subsections we consider the problems of truncating this invariant to lines and quadrangles of the dual Fano plane. Using the black hole analogy
we show that the situation of truncation to a line corresponds to the one of truncating the $N=8$ black hole scenario with moduli space $E_{7(7)}/SU(8)$  to the $N=4$ one with moduli space
$SL(2)/U(1)\times SO(6,6,)/SO(6)\times SO(6)$.
We find that the truncation to a quadrangle complementary to the line     giving rise to the $N=4$ case is just the $N=2$ truncation with the moduli space being $SO^{\ast}(12)/U(6)$. 
Finally an explicit correspondence between the $56$ integer-valued amplitudes
of the seven qubits and the $56$ charges  ($28$ electric and $28$ magnetic) is established.
This result provides a link between the well-known $56$ dimensional  representation of $E_7$
as given by Cartan\cite{Cartan} and the one suggested by the quantum information theoretic analogy.
The conlcusions and the discussion are left for Section VII.
Here we also comment on a possibility of developing a qubit version of the
solution of the attractor equations in the spirit of Ref. 5. in the more general $N=8$ context.
This process should correspond to the optimization of the entanglement
distributed along the entangled lattice defined by the Fano plane.
In closing we conjecture that using the techniques as developed in this paper and the similar ones developed in the mathematical literature by Elduque\cite{Elduque} and Manivel\cite{Manivel}
a quantum information theoretic approach also to magic supergravities might exist. 
These approaches might provide additional insight into the structure of such theories.

\section{The group $SO(4,4,\bf{C})$ and four qubits}

A four qubit state can be written in the form

\beq
\vert {\Psi}\rangle =\sum_{i_1i_2i_3i_4=0,1}{\Psi}^{i_1i_2i_3i_4}\vert i_1i_2i_3i_4\rangle,\quad \vert i_1i_2i_3i_4\rangle\equiv \vert i_1\rangle\otimes\vert i_2\rangle\otimes\vert i_3\rangle\otimes\vert i_4\rangle\in V_1\otimes V_2\otimes V_3
\otimes V_4,
\eeq
\noindent
where $V_a\equiv {\bf C}^2$ , $a=1,2,3,4$.
In this notation $\vert i_a\rangle$ ($i_a=0,1$) are the canonical basis vectors of the $a$th qubit.
The group of stochastic local operations and classical communication\cite{Vidal}(SLOCC)
representing admissible fourpartite protocols is $SL(2, {\bf C})^{\otimes 4}$ acting on $\vert\Psi\rangle$ as

\beq
\vert\Psi\rangle\mapsto S_1\otimes S_2\otimes S_3\otimes S_4\vert\Psi\rangle,\quad S_a\in SL(2, {\bf C}),\quad a=1,2,3,4,
\eeq
\noindent
where the label $a$ refers to the qubit the SLOCC transformation is acting on.

Our aim in this subsection is to give a unified description of four-qubit states {\it and} SLOCC transformations. As we will see states and transformations taken together
can be described in a unified manner using the group $SO(4,4,{\bf C})$.
This point of view is based on the idea of a dual characterization of four-qubits as states and at the same time as transformations. Entangled states representing configurations of quantum entanglement in this picture are also regarded as {\it operators}. An entangled state is a pattern of entanglement, however this pattern of entanglement can also be regarded as a one acting on other patterns of entanglement to produce new kind of entanglement. 
This situation is reminiscent of the situation one encounters in topological field theory, braid groups etc. where there is a shift from elements of a topological category to morphisms in an associated category and vice versa.
Let us see how this works in the case of four qubits.

We can arrange the $16$ complex amplitudes appearing in ${\Psi}^{i_1i_2i_3i_4}$
in a $4\times 4 $ matrix in six different ways

\beq
D_0(\vert\Psi\rangle)=\begin {pmatrix}{\Psi}^{0000}&{\Psi}^{0001}&{\psi}^{0010}&{\Psi}^{0011}\\
{\Psi}^{0100}&{\Psi}^{0101}&{\Psi}^{0110}&{\Psi}^{0111}\\
{\Psi}^{1000}&{\Psi}^{1001}&{\Psi}^{1010}&{\Psi}^{1011}\\
{\Psi}^{1100}&{\Psi}^{1101}&{\Psi}^{1110}&{\Psi}^{1111}\end {pmatrix},
\eeq
\noindent

\beq
D_1(\vert\Psi\rangle)=\begin {pmatrix}{\Psi}^{0000}&{\Psi}^{0001}&{\psi}^{0100}&{\Psi}^{0101}\\
{\Psi}^{0010}&{\Psi}^{0011}&{\Psi}^{0110}&{\Psi}^{0111}\\
{\Psi}^{1000}&{\Psi}^{1001}&{\Psi}^{1100}&{\Psi}^{1101}\\
{\Psi}^{1010}&{\Psi}^{1011}&{\Psi}^{1110}&{\Psi}^{1111}\end {pmatrix},
\eeq
\noindent

\beq
D_2(\vert\Psi\rangle)=\begin {pmatrix}{\Psi}^{0000}&{\Psi}^{0001}&{\psi}^{1000}&{\Psi}^{1001}\\
{\Psi}^{0010}&{\Psi}^{0011}&{\Psi}^{1010}&{\Psi}^{1011}\\
{\Psi}^{0100}&{\Psi}^{0101}&{\Psi}^{1100}&{\Psi}^{1101}\\
{\Psi}^{0110}&{\Psi}^{0111}&{\Psi}^{1110}&{\Psi}^{1111}
\end {pmatrix}.
\eeq
\noindent
and their transposed  matrices $D_0^T$, $D_1^T$ and $D_2^T$.

Let us introduce the $2\times 2$ matrices $E_{ij}$, $i,j=0,1$ as

\beq
E_{00}=\begin {pmatrix}1&0\\0&0\end {pmatrix},\quad
E_{01}=\begin {pmatrix}0&1\\0&0\end {pmatrix},\quad
E_{10}=\begin {pmatrix}0&0\\1&0\end {pmatrix},\quad
E_{11}=\begin {pmatrix}0&0\\0&1\end {pmatrix}.
\eeq
\noindent
Then it is straightforward to check that the matrices above can be written as

\beq
D_0(\vert\Psi\rangle)={\Psi}^{i_1i_2i_3i_4}E_{i_1i_3}\otimes E_{i_2i_4}: V_3\otimes V_4\to V_1\otimes V_2.
\eeq
\beq
D_1(\vert\Psi\rangle)={\Psi}^{i_1i_2i_3i_4}E_{i_1i_2}\otimes E_{i_3i_4}: V_2\otimes V_4\to V_1
\otimes V_3.
\eeq
\beq
D_2(\vert\Psi\rangle)={\Psi}^{i_1i_2i_3i_4}E_{i_2i_1}\otimes E_{i_3i_4}: V_1\otimes V_4\to V_2
\otimes V_3.
\eeq
\noindent
and the transposed maps $D_0^T$, $D_1^T$ and $D_2^T$ map the spaces in reversed order.
Hence we see that a four-qubit state can also be regarded as an operator mapping a two-qubit state which is an element of $V_a\otimes V_b$ to another one belonging to $V_c\otimes V_d$.  

Let us define now the natural symplectic structure                              $\omega: V\times V\to {\bf C}$ on $V\equiv {\bf C}^2$ given by its action on the computational base

\beq
\omega(\vert i\rangle,\vert j\rangle)\equiv {\varepsilon}_{ij}=
\begin {pmatrix}0&1\\-1&0\end {pmatrix}.
\eeq
\noindent
Define moreover, the new matrices

\beq
{\cal D}_0(\vert\Psi\rangle)={\Psi}^{i_1i_2i_3i_4}E_{i_1i_3}\varepsilon\otimes E_{i_2i_4}\varepsilon.
\eeq
\noindent
\beq
{\cal D}_1(\vert\Psi\rangle)={\Psi}^{i_1i_2i_3i_4}E_{i_1i_2}\varepsilon\otimes E
_{i_3i_4}\varepsilon.
\eeq
\noindent
\beq
{\cal D}_2(\vert\Psi\rangle)={\Psi}^{i_1i_2i_3i_4}E_{i_2i_1}\varepsilon\otimes E
_{i_3i_4}\varepsilon.
\eeq
\noindent

First we construct three $8\times 8$ representations  of $SO(4,4,{\bf C})$ as follows (triality).
The carrier of the $8$ dimensional representation ${\cal R}_0$
is the space $V_1\otimes V_2\oplus V_3\otimes V_4$. The carrier spaces 
of the representations ${\cal R}_1$ and ${\cal R}_2$ are obtained by permuting the $123$ labels of the vector spaces $V_{1,2,3}$ accordingly,
i.e. they are $V_1\otimes V_3\oplus V_2\otimes V_4$ and $V_2\otimes V_3\oplus V_1\otimes V_4$.
Let us define the Wootters\cite{Wootters} spin flip operation used in quantum information
as

\beq
\tilde{\cal D}_I\equiv ({\sigma}_2\otimes {\sigma}_2){\cal D}_I^T({\sigma}_2\otimes {\sigma}_2)=({\varepsilon}\otimes {\varepsilon}){\cal D}_I^T({\varepsilon}\otimes {\varepsilon}),\quad I=0,1,2.
\eeq
\noindent
Then the three representatives of the four-qubit states regarded now as suitable operators intertwining pairs of qubits (related to the $8_v-8_s-8_c$ triality) 
can be written in the form

\beq
{\cal R}_I(\vert\Psi\rangle)=\begin {pmatrix}0&{\cal D}_I(\vert\Psi\rangle)\\
-\tilde{\cal D}_I(\vert\Psi\rangle)&0\end {pmatrix},\quad I=0,1,2.
\eeq
\noindent

Hence using the triality construction above we can associate to the $16$ basis vectors of the four-qubit Hilbert space $3\times 16$,  $8\times 8$ matrices as follows

\beq
{\cal R}_0(\vert i_1i_2i_3i_4\rangle)=\begin {pmatrix}0&E_{i_1i_3}\varepsilon\otimes E_{i_2i_4}\varepsilon\\-E_{i_3i_1}\varepsilon\otimes E_{i_4i_2}\varepsilon&0\end {pmatrix} \quad {\rm on}\quad V_1\otimes V_2\oplus V_3\otimes V_4
\eeq 
\noindent

\beq
{\cal R}_1(\vert i_1i_2i_3i_4\rangle)=\begin {pmatrix}0&E_{i_1i_2}\varepsilon
\otimes E_{i_3i_4}\varepsilon\\-E_{i_2i_1}\varepsilon\otimes E_{i_4i_3}\varepsilon&
0\end {pmatrix} \quad {\rm on} \quad V_1\otimes V_3\oplus V_2\otimes V_4
\eeq
\noindent

\beq
{\cal R}_2(\vert i_1i_2i_3i_4\rangle)=\begin {pmatrix}0&E_{i_2i_1}\varepsilon 
\otimes E_{i_3i_4}\varepsilon\\-E_{i_1i_2}\varepsilon\otimes E_{i_4i_3}\varepsilon&
0\end {pmatrix}\quad {\rm on}\quad  V_2\otimes V_3\oplus V_1\otimes V_4.
\eeq
\noindent

How can we also include the SLOCC transformations in this $SO(4,4,{\bf C})$ picture?  
Since $SL(2, {\bf C})^{\otimes 4}\subset SO(4,4,{\bf C})$ it is natural to expect that the generators of its Lie algebra $(s_1,s_2,s_3,s_4)\in sl(2,{\bf C})\oplus sl(2, {\bf C})\oplus sl(2, {\bf C})\oplus sl(2, {\bf C})\equiv sl(2,{\bf C})^{\oplus 4}$
are represented by suitably defined block diagonal entries of the representations ${\cal R}_I$.
In order to see that it is really the case we should define in the qubit picture a convenient 
realization for the generators of $sl(2, {\bf C})$ .

Define the matrices

\beq
s_{ij}\equiv\frac{1}{2}(E_{ij}\varepsilon+E_{ji}\varepsilon),
\eeq
\noindent

i.e. these are of the form

\beq
s_{00}=\begin {pmatrix}0&1\\0&0\end {pmatrix},\quad
s_{11}=\begin {pmatrix}0&0\\-1&0\end {pmatrix},\quad
s_{01}=s_{10}=\frac{1}{2}\begin {pmatrix}-1&0\\0&1\end {pmatrix}.
\eeq
\noindent
Let us define the representatives of these generators of $sl(2, {\bf C})^{\oplus 4}$ as follows

\beq
{\cal R}_0(s_{i_1j_1},s_{i_2j_2},s_{i_3j_3},s_{i_4j_4})=
\begin {pmatrix}s_{i_1j_1}\otimes I+I\otimes s_{i_2j_2}&0\\
0&s_{i_3j_3}\otimes I+I\otimes s_{i_4j_4}\end {pmatrix}
\eeq
\noindent

\beq
{\cal R}_1(s_{i_1j_1},s_{i_2j_2},s_{i_3j_3},s_{i_4j_4})=
\begin {pmatrix}s_{i_1j_1}\otimes I+I\otimes s_{i_3j_3}&0\\
0&s_{i_2j_2}\otimes I+I\otimes s_{i_4j_4}\end {pmatrix}
\eeq
\noindent

\beq
{\cal R}_0(s_{i_1j_1},s_{i_2j_2},s_{i_3j_3},s_{i_4j_4})=
\begin {pmatrix}s_{i_2j_2}\otimes I+I\otimes s_{i_3j_3}&0\\
0&s_{i_1j_1}\otimes I+I\otimes s_{i_4j_4}\end {pmatrix}.
\eeq
\noindent

In order to check that these $3\times 4\oplus 16=28$ generators close to form a Lie-algebra which is just $so(4,4,{\bf C})$ the only commutator we have to check
is the one 

\beq
[{\cal R}_I(\vert i_1i_2i_3i_4\rangle),{\cal R}_I(\vert j_1j_2j_3j_4\rangle)].
\eeq
\noindent
Indeed, the remaining ones are trivial due to the subalgebra property
of $sl(2,{\bf C})^{\oplus 4}$, and the commutator ${\cal R}_I([(s_1,s_2,s_3,s_4),\vert i_1i_2i_3i_4\rangle])$ is defined by the very action of the SLOCC group on the four qubit states.
To calculate the commutator Eq. (24) we need the identities

\beq
E_{ij}\varepsilon -E_{ji}\varepsilon={\varepsilon}_{ji} I
\eeq
\noindent
\beq
(E_{ij}\varepsilon)(E_{kl}\varepsilon)={\varepsilon}_{jk}(E_{il}\varepsilon),
\eeq
\noindent
which can be proved using the identity $\omega (v,u)w+\omega(w,v)u+\omega(u,w)v=0$ for $v=\vert i\rangle, u=\vert j\rangle, w=\vert k\rangle\in V$,
and the fact that the matrix representation of the linear map $\vert i\rangle\omega (\vert j\rangle, \cdot): V\to V$ is just $E_{ij}\varepsilon$. Finally ,we obtain for the commutator Eq. (24) the form

\beq
[{\cal R}_I(\vert i_1i_2i_3i_4\rangle),{\cal R}_I(\vert j_1j_2j_3j_4\rangle)]=
\sum_{a=1}^4\left(\prod_{a\neq b}{\varepsilon}_{i_bj_b}\right){\cal R}_I(s_{i_aj_a})
,
\eeq
\noindent
here we slightly simplified the notation
by defining e.g. ${\cal R}(s_{i_3j_3})\equiv {\cal R}(0,0,s_{i_3j_3},0)$ etc.
Hence we have a closed Lie-algebra structure.
In order to show that it is just the algebra $so(4,4,{\bf C})$ the only thing we have to do is to use Eqs. (14-15) to show that

\beq
{\cal R}_IG+G{\cal R}_I^T=0,\quad {\rm with}\quad
G\equiv\begin {pmatrix}{\varepsilon}\otimes {\varepsilon}&0\\0&{\varepsilon}\otimes{\varepsilon}\end {pmatrix}\quad I=0,1,2
\eeq
\noindent
which indeed holds.

\section{The group $E_7({\bf C})$ and the Fano plane}

In the paper of Duff and Ferrara seven qubits were considered which are entangled in a very special way. The seven qubits contain only tripartite entanglement, and the distribution of this curious type of entanglement is governed by the geometry of the second largest exceptional group $E_7({\bf C})$.
Explicitly Duff and Ferrara considers the state

\begin{eqnarray}
\vert\psi\rangle&=& a_{ABC}\vert ABC\rangle +b_{CEF}\vert CEF\rangle+
c_{BFD}\vert BFD\rangle +d_{DAE}\vert DAE\rangle\nonumber\\
&+&e_{EBG}\vert EBG\rangle +f_{FGA}\vert FGA\rangle +g_{GDC}\vert GDC\rangle,
\end{eqnarray}
\noindent
where in their notation $A,B,C,D,E,F,G=0,1$ and these labels are corresponding
to seven parties: Alice, Bob, Charlie, Daisy, Emma, Fred and George.
(Note that for our later convenience we slightly changed their notation by exchanging $B$ and $C$.) 
As the authors notice the entanglement encoded in the state ${\vert\Psi\rangle}$ has the following properties:

{\bf 1.} Any pair of states has an individual in common.

{\bf 2.} Each individual is excluded from four out of the seven states.

{\bf 3.} Two given individuals are excluded from two out of the seven states.

{\bf 4.} Three given individuals never excluded.

Then they represent this type of seven qubit entanglement by a heptagon with vertices $A,B,C,D,E,F,G$ and triangles $ABC$, $CEF$, $BFD$, $DAE$, $EBG$, $FGA$ and $GDC$.

It is easy to show that the properties ${\bf 1.}-{\bf 4.}$ together with the heptagon picture is equivalent to attaching seven qubits to the vertices of the Fano plane. The Fano plane, the smallest projective plane, (see Fig. 1.) is a little gadget containing seven points and seven lines. 
                                                                                \begin{figure}                                                                  \centerline{\resizebox{9.0cm}{!}{\includegraphics{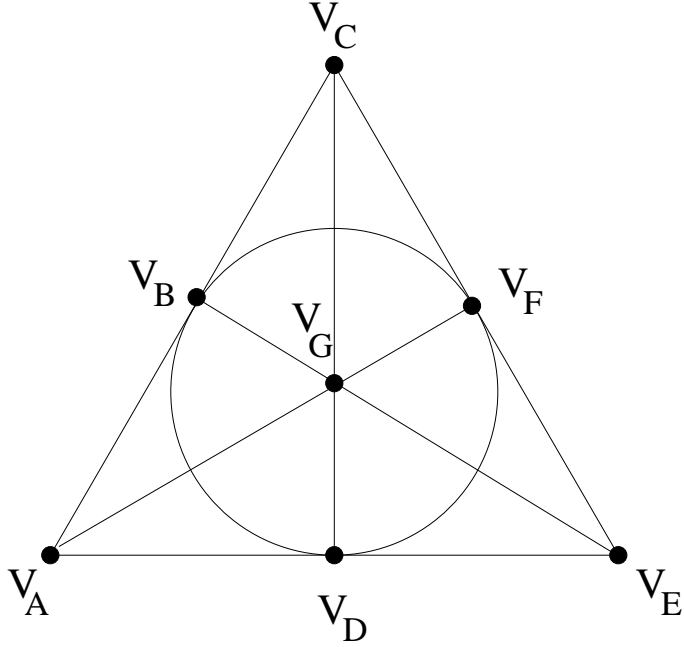}}}                 \caption{\label{figr1}
The Fano plane with seven copies of the two dimensional complex vector space $V$ corresponding to qubits $A, B, C, D, E, F,$ and $G$ attached to its vertices. 
} \end{figure} 
It has three points on every line and three lines through every point.
The two dimensional complex vector spaces corresponding to the seven qubits of the seven parties are denoted by $V_A$, $V_B$, $V_C$, $V_D$, $V_E$, $V_F$ and $V_G$. These vector spaces are attached to the vertices of the Fano plane.
The seven projective lines in this picture correspond to the seven triangles of Duff and Ferrara representing tripartite entanglement.
Looking at Fig. 1. it is obvious that properties ${\bf 1.}-{\bf 4.}$ are satisfied. Moreover, the incidence graph of the Fano plane can be described by using two heptagons with their vertices corresponding to lines and points respectively.
An alternative picture is obtained by using  merely one heptagon with seven triangles representing the lines. This is precisely what can be seen in Fig. 1. of Duff and Ferrara. 

Let us now adopt a ${\bf Z}_2^3$ labelling of our qubits (points in the Fano plane),

\beq
(A,B,C,D,E,F,G)\mapsto (1,2,3,4,5,6,7)\mapsto (001,010,011,100,101,110,111),
\eeq
\noindent
see Fig.2.
\begin{figure}
\centerline{\resizebox{9.0cm}{!}{\includegraphics{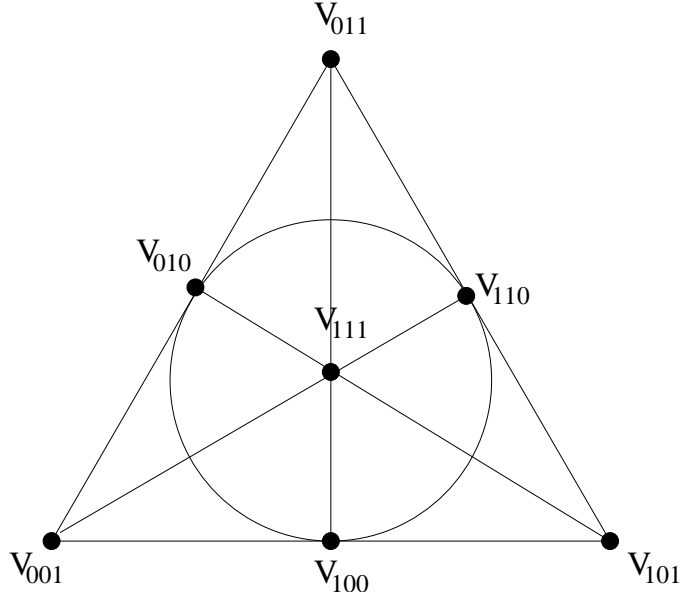}}}
\caption{\label{figr3}
The Fano plane with the associated qubits labelled by elements $\sigma$ of ${\bf Z}_2^3$.
 } \end{figure}
For the labelling of the tripartite states (lines of the Fano plane) we choose 
the convention (Fig. 3.)

\begin{figure}
\centerline{\resizebox{9.0cm}{!}{\includegraphics{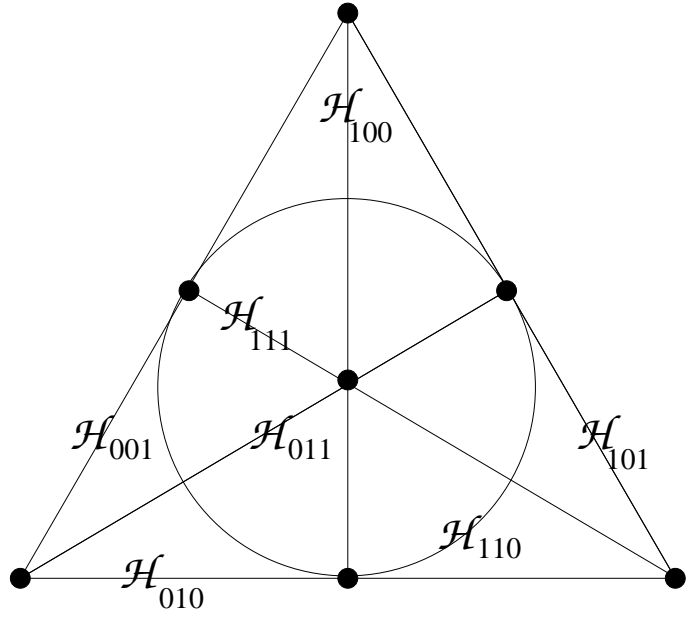}}}      
\caption{\label{figr2}
The Fano plane with the tripartite spaces ${\cal H}_{\sigma}$ , $\sigma\in{\bf Z}_2^3$  associated to its lines.
 } \end{figure}

\beq
{\cal H}_{001}=
V_1\otimes V_2\otimes V_3=V_{001}\otimes V_{010}\otimes V_{011}
\eeq                                                                            \beq
{\cal H}_{010}=V_1\otimes V_4\otimes V_5=V_{001}\otimes V_{100}\otimes V_{101}
\eeq
\beq
{\cal H}_{011}=V_1\otimes V_6\otimes V_7=V_{001}\otimes V_{110}\otimes V_{111}
\eeq
\beq
{\cal H}_{100}=V_3\otimes V_4\otimes V_7=V_{011}\otimes V_{100}\otimes V_{111}
\eeq
\beq
{\cal H}_{101}=V_3\otimes V_5\otimes V_6=V_{011}\otimes V_{101}\otimes V_{110}
\eeq
\beq
{\cal H}_{110}=V_2\otimes V_4\otimes V_6=V_{010}\otimes V_{100}\otimes V_{110}
\eeq
\beq
{\cal H}_{111}=V_2\otimes V_5\otimes V_7=V_{010}\otimes V_{101}\otimes V_{111}.
\eeq
We can neatly summarize this labelling convention by using the ${\bf Z}_2$ linear map 
\begin {eqnarray}
{\chi}: {\bf Z}_2^3&\to& {\bf Z}_2^7\nonumber \\
 (\alpha,\beta,\gamma)&\mapsto& (\alpha,\alpha +\beta,\beta,\gamma,\alpha +\gamma,\alpha +\beta +\gamma,\beta +\gamma)
\end {eqnarray}
\noindent
For example $\chi$ maps $(110)$ to $(1010101)$. The zeros are at the $2$nd,
$4$th and $6$th slot, so the tensor product structure of the sixth tripartite system i.e. ${\cal H}_{110}$ is $V_2\otimes V_4\otimes V_6=V_B\otimes V_D\otimes V_F$ i.e. the three-qubit state of Bob, Daisy and Fred.
Notice that the $7\times 7$ matrix

\beq
\begin {pmatrix} 0&0&0&1&1&1&1\\0&1&1&0&0&1&1\\0&1&1&1&1&0&0\\
1&1&0&0&1&1&0\\1&1&0&1&0&0&1\\1&0&1&0&1&0&1\\1&0&1&1&0&1&0\end {pmatrix}
\eeq
\noindent
is just the complement of the incidence matrix of the Fano plane with this labelling convention.
(If the rows of this matrix label lines and its columns label points, then a particular element of the complement of the incidence matrix is $0$ if the given point belongs to the given line , otherwise it is $1$. For example in the sixth column we have
zeros in the third, fifth and sixth row corresponding to the fact that the sixth point of the Fano plane belongs to third, fifth and sixth line. See Figs 2. and 3.)
Note, that the incidence matrix above is intimately related to error correcting codes, a topic under intense scrutiny in quantum information.
Namely, it is related to the Hamming [7,4] code correcting up to $1$ and detecting up to $3$ errors. Its seven codewords are precisely the rows of this matrix. This code is a maximum distance separable, MDS code\cite{Saniga,Nielsen}.
Later we will see that the incidence matrix encodes all the information
to build up the $56$ dimensional representation of the group $E_7({\bf C})$.
This coincidence might provide another interesting link between 
the topics of stringy black holes and quantum information theory\cite{Duff,Linde,Levay,Ferrara}.

Having clarified the connection between the entangled state $\vert\psi\rangle$
of Eq. (29) and the Fano plane, now we turn to the question of
clarifying its relation to the group $E_7({\bf C})$.
Let us consider the $56$ dimensional complex vector space

\beq
{\cal H}={\cal H}_{001}\oplus {\cal H}_{010}\oplus {\cal H}_{011}\oplus {\cal H}_{100}\oplus{\cal H}_{101}\oplus {\cal H}_{110}\oplus {\cal H}_{111}=\bigoplus_  {\sigma\neq(000)}{\cal H}_{\sigma},\quad \sigma\in{\bf Z}_2^3.
\eeq
\noindent
In this notation the state $\vert\psi\rangle$ of Eq. (29) of seven qubits containing only tripartite entanglement
can be represented as a seven component vector

\beq
\begin {pmatrix} {\psi}^{001}\\{\psi}^{010}\\{\psi}^{011}\\{\psi}^{100}\\{\psi}^{101}\\{\psi}^{110}\\{\psi}^{111}\end {pmatrix}=\begin {pmatrix} a_{ABC}\\d_{ADE}\\f_{AFG}\\g_{CDG}\\b_{CEF}\\c_{BDF}\\e_{BEG}\end {pmatrix}
\eeq
\noindent
where each component in turn has eight components.
Notice that the tensor product structure of the components is entirely fixed by the map ${\chi}$ of Eq. (38), or the incidence matrix Eq. (39).
For example let us consider ${\psi}^{101}$ i.e. the fifth state.
In the fifth row of the incidence matrix we have zeros at the third, fifth and sixth place. Hence we have the tensor product structure                          $V_3\otimes V_5\otimes V_6=V_C\otimes V_E\otimes V_F$. Hence the tripartite subsystem in question is that of Charlie, Emma and Fred.
Alternatively we can label the eight amplitudes of this state as ${\psi}^{101}_{i_3i_5i_6}$.

Let us now motivate the result (to be discussed in the next section) that a natural action of $e_7$ i.e. the Lie-algebra of $E_7$ 
can be defined on ${\cal H}$.
Looking back at Eqs. (31-37) we can see that ${\cal H}$ can also be written in  the following form

\beq
{\cal H}=(V_1\otimes V_2\otimes V_3)\oplus V_1\otimes (V_{45}\oplus V_{67})\oplus
V_2\otimes(V_{46}\oplus V_{57})\oplus V_3\otimes (V_{56}\oplus V_{47}),
\eeq
\noindent
where we introduced the abbreviation $V_{45}\equiv V_4\otimes V_5$ etc.
According to the results of the previous section we can define an action of the generators of $so(4,4,{\bf C})$ regarded as elements of $V_4\otimes V_5\otimes V_6\otimes V_7$ on the spaces $V_{45}\oplus V_{67}$, $V_{46}\oplus V_{57}$ and
 $V_{56}\oplus V_{47}$
consisting of pairs of bipatite states. Indeed the generators acting on those spaces are precisely the ones defined by the representations ${\cal R}_0$, ${\cal R}_1$ and ${\cal R}_2$ (see Eqs. (16-18) and Eqs. (21-23)).
These generators are of two type. The ones defined by the basis vectors of four-qubit states i.e. the ones of ${\cal R}_I(\vert i_4i_5i_6i_7\rangle )$
, and the ones ${\cal R}_I(s_4,s_5,s_6,s_7)$ coming from the generators of the SLOCC group.
Moreover, we see that this action is annihilating the subspace ${\cal H}^{001}=V_1\otimes V_2\otimes V_3$ i.e. the tripartite state of Alice , Bob and Charlie.
Let us label the $so(4,4,{\bf C})$ generators 
${\cal R}_I(\vert i_4i_5i_6i_7\rangle )$
by the label of the subspace they are annihilating.
We define

\beq
{\cal R}_I(T^{001}_{i_4i_5i_6i_7})\equiv{\cal R}_I(\vert i_4i_5i_6i_7\rangle ).
\eeq
\noindent
Notice again that the label $(001)$ defines the  line  $(001)$ on the Fano plane.
Is complement is a quadrangle. In this case this is the quadrangle defined by the points $4567$. Hence the alternative notation ${\cal R}_I(T^{001}_{ijkl})$ uniquely defines the generator in question.
Indeed, the label $(001)$ tells that the spaces involved
are $V_4, V_5, V_6$ and $V_7$ with the corresponding indices $i,j,k$ and $l$ taking the values $0,1$,  and the label $I=0,1,2$ fixes the pairs of bipartite states it is acting on, namely $V_{45}\oplus V_{67}$, $V_{46}\oplus V_{57}$ and $V_{56}\oplus V_{47}$ respectively.
Hence again as in the case of states (Eq.(41)) also in the case of operators it is enough to consider merely the comfortable ${\bf Z}_2^3$ labelling.

Now for the good part. Notice that ${\cal R}_I(T^{001})$ is annihilating ${\cal H}^{001}$ and intertwining the remaining subspaces
as follows
\begin {eqnarray}
{\cal R}_0(T^{001})&:&{\cal H}_{010}\leftrightarrow {\cal H}_{011}\nonumber\\
{\cal R}_1(T^{001})&:&{\cal H}_{110}\leftrightarrow {\cal H}_{111}\\
{\cal R}_2(T^{001})&:&{\cal H}_{100}\leftrightarrow {\cal H}_{101}\nonumber.
\end {eqnarray}
\noindent
Eq.(44) has a nice physical interpretation.
Let us regard the ${\bf Z}_2^3$ labels as some sort of discrete charge labelling inequivalent superselection sectors. Hence the seven tripartite states  of     Eq.(40) formed from the seven qubits have different ${\bf Z}_2^3$ charge.
Moreover, let the operators acting on such states also have such charge.
Then Eq.(44) shows that the particular transformation ${\cal R}(T^{001})$ is connecting those tripartite states for which the ${\bf Z}_2^3$ sum rule $\sigma +{\sigma}^{\prime}={\sigma}^{\prime\prime}$ is satisfied .
The usual four-qubit infinitesimal SLOCC transformations on the qubits of Daisy, Emma, Fred and George are not changing the superselection sector, they are neutral with respect to this ${\bf Z}_2^3$ charge. 
However the ones of Eq.(44) are representing a new type of transformations corresponding to protocols operating between tripartite sectors with different charge.

The next step is to generalize these observations by considering all of our seven tripartite systems equivalent.
We have seven lines corresponding to these systems. The complements of these lines define unique quadrangles in the Fano plane. Since to the points of the Fano plane we have attached qubits, it follows that to its quadrangles we can assign four-qubit states. These four-qubit states according to the correspondence of Section II. define operators acting on pairs of bipartite states.
Hence by introducing six further decompositions of the type as given by Eq. (42) we can construct six further $so(4,4,{\bf C})$ actions on ${\cal H}$.
We can organize the generators of the SLOCC transformations of these seven $so(4,4,{\bf C})$ algebras to a single $sl(2, {\bf C})^{\oplus 7}$ algebra.
These $7\times 3=21$ generators have no ${\bf Z}_2^3$ charge.
However, there are also operators having the ${\bf Z}_2^3$ charges: $(001)$,
$(010)$, $(011)$, $(100)$, $(101)$, $(110)$ and $(111)$.
Their number is $7\times 16=112$.
Hence altogether we have a $133$ complex dimensional vector space, which hopefully can be given the structure of a Lie-algebra, which should be $e_7({\bf C})$.
Luckily some recent results in the mathematics literature guarantee that this construction  indeed yields the Lie-algebra $e_7({\bf C})$. In the next section we attempt a quantum information theoretic motivation of this result. Moreover we also explicitly construct the $56$ dimensional representation of $e_7$ in terms of transformations corresponding to sevenpartite protocols.
As we have already learnt these are of two type: SLOCC transformations, and transformations connecting the different tripartite systems.
It is the construction of these transformations now we turn.

\section{The Lie algebra of $E_7$ in terms of seven qubits}

In the previous section we have conjectured that the Lie-algebra $e_7({\bf C})$ can be understood as a direct sum of eight vector spaces.
The first one is $sl(2, {\bf C})^{\oplus 7}$ and the remaining seven ones are
related to
four-qubit states associated to the seven quadrangles of the Fano plane that are in turn complements to the lines corresponding to our seven tripartite states. 
Let us introduce the notation

\beq
W_{000}=sl(2,{\bf C})^{\oplus 7}.
\eeq
\noindent
\beq
W_{001}=V_4\otimes V_5\otimes V_6\otimes V_7=V_{100}\otimes V_{101}\otimes V_{110}\otimes V_{111}\nonumber
\eeq
\noindent
\beq
W_{010}=V_2\otimes V_3\otimes V_6\otimes V_7=V_{010}\otimes V_{011}\otimes V_{110}\otimes V_{111}\nonumber
\eeq
\noindent
\beq
W_{011}=V_2\otimes V_3\otimes V_4\otimes V_5=V_{010}\otimes V_{011}\otimes V_{100}\otimes V_{101}\nonumber
\eeq
\noindent
\beq
W_{100}=V_1\otimes V_2\otimes V_5\otimes V_6=V_{001}\otimes V_{010}\otimes V_{101}\otimes V_{110}\nonumber
\eeq
\noindent
\beq
W_{101}=V_1\otimes V_2\otimes V_4\otimes V_7=V_{001}\otimes V_{010}\otimes V_{100}\otimes V_{111}\nonumber
\eeq
\noindent
\beq
W_{110}=V_1\otimes V_3\otimes V_5\otimes V_7=V_{001}\otimes V_{011}\otimes V_{101}\otimes V_{111}\nonumber
\eeq
\noindent
\beq
W_{111}=V_1\otimes V_3\otimes V_4\otimes V_6=V_{001}\otimes V_{011}\otimes V_{100}\otimes V_{110}\nonumber
\eeq
Following Ref. 9. let us now define an $sl(2,{\bf C})^{\oplus 7}$ invariant map

\beq
F_{\sigma,\tau}: W_{\sigma}\times W_{\tau}\longrightarrow W_{\sigma+\tau},\quad \sigma,\tau\in {\bf Z}_2^3
\eeq
\noindent
as follows.

{\bf 1.} $F_{(000),(000)}$ is just the Lie bracket in $sl(2, {\bf C})^{\oplus 7}$.

{\bf 2.} $F_{(000),({\sigma}_3{\sigma}_2{\sigma}_1)}=-F_{({\sigma}_3{\sigma}_2{\sigma}_1),(000)}$ is given by the action of $sl(2, {\bf C})^{\oplus 7}$ on $W_{\sigma}$.
For example

\begin {eqnarray}
F_{(000),(001)}((s_1,s_2,s_3,s_4,s_5,s_6,s_7),\vert i_4i_5i_6i_7\rangle)&=&\nonumber\\
(s_4\otimes I\otimes I\otimes I+I\otimes s_5\otimes I\otimes I+
I\otimes I\otimes s_6\otimes I&+& I\otimes I\otimes I\otimes s_7)\vert i_4i_5i_6i_7\rangle.
\end {eqnarray}
\noindent

{\bf 3.} For $F_{\sigma,\sigma}$ we take seven copies of the $so(4,4,{\bf C})$
commutator giving rise to Eq. (27) of Section II. For example

\beq
F_{(101),(101)}(\vert i_1i_2i_4i_7\rangle ,\vert j_1j_2j_4j_7\rangle)=
-\sum_{a=1,2,4,7}\left(\prod_{a\neq b}{\varepsilon}_{i_bj_b}\right)s_{i_aj_a},
\eeq
\noindent
where $b=1,2,4,7$ and we used the abbreviation e.g. $s_{i_2j_2}=(0,s_{i_2j_2},0,0,0,0,0)$. Notice also that here we have also included a minus sign to get nicer expressions later on. 

{\bf 4.} For $\sigma\neq\tau$ and $\sigma\neq (000)\neq \tau$ $F_{\sigma,\tau}$
is obtained by contraction with two ${\varepsilon}$s in the indices corresponding to the two parties the two four-qubit states share. 
(Recall, all of the pairs taken from the seven quadrangles in the Fano plane have two points in common.) For example

\beq
F_{(110),(011)}(\vert i_1i_3i_5i_7\rangle ,\vert j_2j_3j_4j_5\rangle)=
\pm{\varepsilon}_{i_3j_3}{\varepsilon}_{i_5j_5}\vert i_1j_2j_4i_7\rangle.
\eeq
\noindent

We can write Eqs. (48-49) in the more instructive form

\beq
[T^{101}_{i_1i_2i_4i_7}, T^{101}_{j_1j_2j_4j_7}]=-\sum_{a=1,2,4,7}\left(\prod_{a\neq b}{\varepsilon}_{i_bj_b}\right)T^{000}_{i_aj_a},
\eeq
\noindent
\beq
[T^{110}_{i_1i_3i_5i_7},T^{011}_{j_2j_3j_4j_5}]=\pm{\varepsilon}_{i_3j_3}{\varepsilon}_{i_5j_5}T^{101}_{i_1j_2j_4i_7},
\eeq
\noindent
by introducing the infinitesimal operators $T^{\sigma}$,  $\sigma \in{\bf Z}_2^3$. The important subtlety here is the calculation of the signs appearing in Eq. (50-51). It was shown in Ref.9. that if we take for the commutators of such type the form

\beq
[T^{\sigma},T^{\tau}]=\phi(\sigma,\tau)F_{\sigma,\tau}(T^{\sigma},T^{\tau}),
\eeq
\noindent
where the value of $\phi(\sigma,\tau)$ is $\pm 1$ and the signs are precisely the ones appearing in the multiplication table of the octonions, then the Lie algebra we get is $e_7(\bf C)$.
Hence we have

\beq
e_7=\bigoplus_{\sigma\in {\bf Z}_2^3} W_{\sigma}.
\eeq
\noindent
The octonionic multiplication table producing the correct signs for $\phi(\sigma,\tau)$ is given in terms of the basis vectors $(f_{000},f_{001},f_{010}, f_{011},f_{100},f_{101},f_{110},f_{111})\equiv (1,i,j,k,l,il,jl,kl)$
as follows.
Using projective duality let us introduce the dual Fano plane with its points corresponding to the lines of the original one. Since to the lines we have attached the tripartite spaces ${\cal H}_{\sigma}$, $\sigma\in {\bf Z}_2^3-(000)$
there is a natural correspondence between these tripartite spaces ${\cal H}_{\sigma}$ and the octonionic basis vectors $f_{\sigma}$, $\sigma\in {\bf Z}_2^3-(000)$.
Moreover, there is also a one-to-one correspondence between lines and their complements (quadrangles) which are related to the superselection sectors $W_{\sigma}$, $\sigma\in {\bf Z}_2^3-(000)$.
Hence it is natural to conjecture that the convention for the signs appearing in the octonionic multiplication table for the basis vectors $f_{\sigma}$ is somehow related to the one appearing in  the commutators of the generators belonging to the sectors $W_{\sigma}$.
In Ref. 9. it has been proved that the sign convention dictated by the octonionic basis $f_{\sigma}$, $\sigma\in{\bf Z}_2^3$ is precisely the one appearing
in the $e_7({\bf C})$  commutators of Eq. (52).
In other words we have an octonionic grading\cite{Manivel}  on $e_7({\bf C})$.

In order to fix our convention concerning the multiplication table of the       octonions
we take an oriented copy of the Fano plane with its points corresponding now to
the octonionic basis vectors labelled as $e_{\sigma}$, $\sigma\in{\bf Z}_2^3$   see Figure 4.
\begin{figure}
\centerline{\resizebox{9.0cm}{!}{\includegraphics{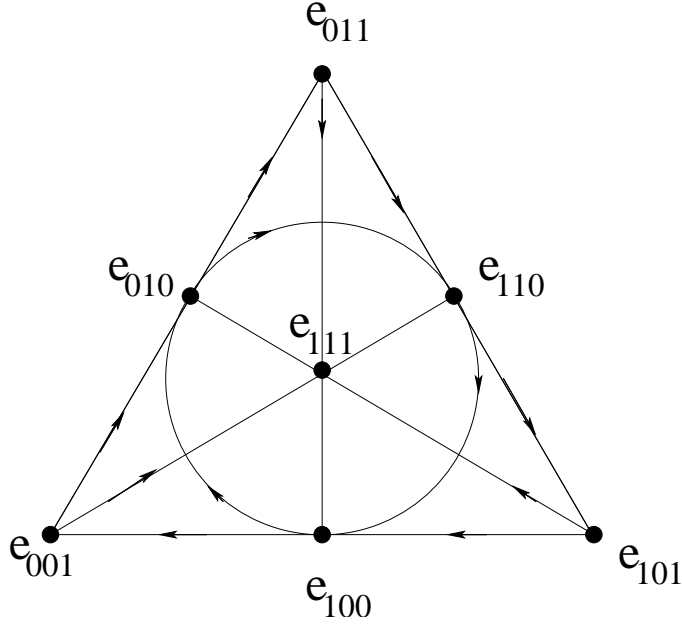}}}
\caption{\label{figr4}
The multiplication table of the octonions as represented by the oriented Fano plane.
 } \end{figure}
The correspondence between the basis vectors of Elduque\cite{Elduque}  ($i$, $j$
, $k$, $l$, $il$, $jl$, $kl$ ) and us is given as follows
\beq
(e_{000}, e_{001}, e_{010}, e_{011},e_{100}, e_{101}, e_{110}, e_{111})          \leftrightarrow
(1, i,j,k,l,-il,-jl,-kl).
\eeq
\noindent
(Note, that in particular $e_{001}e_{100}=-e_{101}$, but $f_{001}f_{100}=f_{101}$.)
Hence from the multiplication table as given by this oriented Fano plane and the correspondence of Eq. (54) $\phi({\sigma,\tau})$ can be obtained.

In closing this section we remark that there is also an explicit formula for $\phi(\sigma, \tau)$. 
It is given by

\beq
\phi(\sigma, \tau)=(-1)^{{\varphi}(\sigma,\tau)},\quad \varphi(({\sigma}_3{\sigma}_2{\sigma}_1),({\tau}_3{\tau}_2{\tau}_1))=\sum_{m\geq n}{\sigma}_m{\tau}_n+{\sigma}_1{\tau}_2{\tau}_3+{\tau}_1{\sigma}_2{\tau}_3+{\tau}_1{\tau}_2{\sigma}_3.
\eeq
\noindent
The existence of this formula is related to the fact that octonions form a twisted group algebra\cite{Majid} satisfying $f_{\sigma}f_{\tau}=\phi(\sigma,\tau)f_{\sigma+\tau}$.
Notice that in particular $\phi(\sigma,\sigma)=-1$, for all $\sigma\in{\bf Z}_2^3-(000)$ giving an explanation for our introducing the extra negative sign in Eq. (48).

\section{The fundamental of $E_7$ in terms of seven qubits}

In order to construct the action of the generators of $e_7$ on the space ${\cal H}$ as a first step
let 
us try to understand the $so(4,4,{\bf C})$ action of one of the generators $T^{\sigma}, \sigma\neq (000)$ e.g. $T^{001}$.  
This operator is associated with the quadrangle $4567$ hence its index structure is $T^{001}_{i_4i_5i_6i_7}$.
First of all let us denote the 
$56$ dimensional representation of $T^{001}$ as $\Sigma (T^{001})$ (leaving the indices $i_4i_5i_6i_7$ implicit).
Then using the decomposition as given by Eq. (42) we have 

\beq
\Sigma(T^{001})=\Delta(T^{001})\Pi^{001}
\eeq
\noindent
where 
\beq
{\Delta}(T^{001})=
\begin {pmatrix} 0&0&0&0&0&0&0\\
0&I\otimes {\cal D}_0&0&0&0&0&0\\
0&0&-I\otimes\tilde{\cal D}_0&0&0&0&0\\
0&0&0&-I\otimes \tilde{\cal D}_2&0&0&0\\
0&0&0&0&I\otimes{\cal D}_2&0&0\\
0&0&0&0&0&I\otimes{\cal D}_1&0\\
0&0&0&0&0&0&-I\otimes\tilde{\cal D}_1
\end {pmatrix}
\eeq
\noindent
and
\beq
{\Pi}^{001}=\begin {pmatrix} {\bf 1}&0&0&0&0&0&0\\
0&0&{\bf 1}&0&0&0&0\\
0&{\bf 1}&0&0&0&0&0\\
0&0&0&0&{\bf 1}&0&0\\
0&0&0&{\bf 1}&0&0&0\\
0&0&0&0&0&0&{\bf 1}\\
0&0&0&0&0&{\bf 1}&0
\end {pmatrix}
\eeq
\noindent
where ${\bf 1}=I\otimes I\otimes I$.
Hence the representation is factorized to a block diagonal matrix (implicitly depending also on the label $ijkl$) and on a permutation (which is only depending on the ${\bf Z_2^3}$ label).
Notice also that the structure of ${\Pi}^{001}$ is entirely fixed by the action of the label $(001)$ on the ${\bf Z}_2^3$ labels of ${\cal H}_{\sigma}$.

Let us now take an operator belonging to a different superselection sector e.g. $T^{010}$.
This operator is associated with the quadrangle $2367$ hence its index structure is
$T^{010}_{j_2j_3j_6j_7}$
Consider now the new decomposition 

\beq
(V_1\otimes V_4\otimes V_5)\oplus V_1\otimes (V_{23}\oplus V_{67})\oplus V_4\otimes (V_{26}\oplus V_{37})\oplus V_5\otimes (V_{36}\oplus V_{27}).
\eeq
\noindent
This vector space is the same as ${\cal H}$ up to ordering of the summands, and the last four terms in this direct sum has been subjected to the permutation $P_{12}$ exchanging the corresponding first two qubits.
From this decomposition it is obvious that now the subspace ${\cal H}_{010}=V_1\otimes V_4\otimes V_5$ is annihilated. Moreover ${\cal R}_0$ is acting on $V_{23}\oplus V_{67}$, ${\cal R}_1$ on $V_{26}\oplus V_{37}$ and ${\cal R}_2$ on
$V_{36}\oplus V_{27}$.
Using this it is easy to see that the operator ${\Sigma}(T^{010})$ acting now on the original space ${\cal H}$ is again of the form ${\Sigma}(T^{010})=\Delta(T^{010}){\Pi}^{010}$ where

\beq
{\Delta}(T^{010})=
\begin {pmatrix} I\otimes {\cal D}_0&0&0&0&0&0&0\\
0&0&0&0&0&0&0\\
0&0&-I\otimes\tilde{\cal D}_0&0&0&0&0\\
0&0&0&P_{12}I\otimes \tilde{\cal D}_1P_{12}&0&0&0\\
0&0&0&0&P_{12}I\otimes{\cal D}_2P_{12}&0&0\\
0&0&0&0&0&-P_{12}I\otimes{\cal D}_1P_{12}&0\\
0&0&0&0&0&0&-P_{12}I\otimes\tilde{\cal D}_2P_{12}
\end {pmatrix}
\eeq
\noindent
and
\beq
{\Pi}^{010}=\begin {pmatrix} 0&0&{\bf 1}&0&0&0&0\\
0&{\bf 1}&0&0&0&0&0\\
{\bf 1}&0&0&0&0&0&0\\
0&0&0&0&0&{\bf 1}&0\\
0&0&0&0&0&0&{\bf 1}\\
0&0&0&{\bf 1}&0&0&0\\
0&0&0&0&{\bf 1}&0&0
\end {pmatrix}
\eeq
\noindent
Using Eq. (12) it is important to realize that for example

\beq
P_{12}\left(I\otimes {\cal D}_1(T^{010}_{j_2j_3j_6j_7})\right)P_{12}=\left(E_{j_2j_3}\right){\varepsilon}\otimes I\otimes \left(E_{j_6j_7}\right){\varepsilon},
\eeq
\noindent
etc. so we expect that the operators occurring in the diagonal of any ${\Sigma}(T^{\sigma})$ are always of the form $I\otimes E\varepsilon\otimes E\varepsilon$,
$E\varepsilon\otimes I\otimes E\varepsilon$ and $E\varepsilon\otimes E\varepsilon\otimes I$. A straightforward check for the remaining five cases shows that this is indeed the case. 
Notice that the explicit form of these operators for a fixed ${\bf Z}_2^3$ label $\sigma$ together with their implicit index structure is controlled by the permutation matrix ${\Pi}^{\sigma}$ and the corresponding decomposition similar to the one in Eq.(42). However, the structure of these ingredients is in turn encoded in the geometry of the Fano plane. 
Hence we can conclude that the structure of the fundamental $56$ dimensional representation of $e_7({\bf C})$ together with the explicit action of its generators can be built from operators representing tripartite protocols on seven tripartite states on an entangled lattice defined by the Fano plane.
The generators of $e_7$ {\it not belonging} to the SLOCC subalgebra $sl(2, {\bf C})^{\oplus 7}$ are always of the form

\beq
\Sigma(T^{\sigma})=\Delta(T^{\sigma}){\Pi}^{\sigma},
\eeq
\noindent
where ${\Delta}$ is a $7\times 7$ diagonal matrix containing a zero and six operators of the form

\beq
(\varepsilon\otimes E\otimes E)(\varepsilon\otimes \varepsilon\otimes \varepsilon),\quad
(E\otimes\varepsilon\otimes E)(\varepsilon\otimes \varepsilon\otimes \varepsilon),\quad
(E\otimes E\otimes\varepsilon)(\varepsilon\otimes \varepsilon\otimes \varepsilon),
\eeq
\noindent
multiplied by a permutation matrix ${\Pi}^{\sigma}$ entangling the different tripartite states.
The action of the infinitesimal SLOCC transformations $sl (2, {\bf C})^{\oplus 7}$  is the usual one, i.e. each of the seven tripartite states transforms under the corresponding three copies of the $2\times 2$ representation of $sl(2,{\bf C})$ and  are behaving as singlets under the remaining four.  

In order to check one set of commutation relations, namely the one $[\Sigma(T^{001}),\Sigma(T^{010})]$ let us also give explicitly the generators of the superselection sector $011$.
Using the decomposition 

\beq
(V_1\otimes V_6\otimes V_7)\oplus V_1\otimes (V_{23}\oplus V_{45})\oplus V_6\otimes (V_{24}\oplus V_{35})\oplus V_7\otimes (V_{34}\oplus V_{25}).
\eeq
\noindent

one can see that they are of the form

\beq
{\Delta}(T^{011})=
\begin {pmatrix} I\otimes {\cal D}_0&0&0&0&0&0&0\\
0&-I\otimes \tilde{\cal D}_0&0&0&0&0&0\\
0&0&0&0&0&0&0\\
0&0&0&{\cal D}_2\otimes I&0&0&0\\
0&0&0&0&-\tilde{\cal D}_1\otimes I&0&0\\
0&0&0&0&0&{\cal D}_1\otimes I&0\\
0&0&0&0&0&0&-\tilde{\cal D}_2\otimes I
\end {pmatrix},
\eeq
\noindent

\beq
{\Pi}^{011}=\begin {pmatrix} 0&{\bf 1}&0&0&0&0&0\\
{\bf 1}&0&0&0&0&0&0\\
0&0&{\bf 1}&0&
0&0&0\\
0&0&0&0&0&0&{\bf 1}\\
0&0&0&0&0&{\bf 1}&0\\
0&0&0&0&{\bf 1}&0&0\\
0&0&0&{\bf 1}&0&0&0
\end {pmatrix}.
\eeq
\noindent
A straightforward calculation using the identities Eqs. (25-26) gives the result

\beq
[\Sigma(T^{001}_{i_4i_5i_6i_7}),\Sigma(T^{010}_{j_2j_3j_6j_7})]={\varepsilon}_{i_6j_6}{\varepsilon}_{i_7j_7}\Sigma(T^{011}_{j_2j_3i_4i_5})
\eeq
\noindent
in accordance with Eq. (52). (Note that according to the octonionic multiplication table 
we have $\phi(001,010)=+1$ hence we have a plus sign.)

Now we have the 56 dimensional matrix representations ${\Sigma}(T^{001}_{i_4i_5i_6i_7})$, ${\Sigma}(T^{010}_{i_2i_3i_6i_7})$ and ${\Sigma}(T^{011}_{i_2i_3i_4i_5})$ of the $e_7({\bf C})$  generators $T^{\sigma}$ for $\sigma=(001),(010),(011)$.   
Clearly these  $48$ operators $\Sigma(T^{\sigma})$ and the $18$ SLOCC generators
are forming the vector space

\beq
sl(2, {\bf C})^{\oplus 6}\otimes W_{001}\oplus W_{010}\oplus W_{011}.
\eeq
\noindent
According to Eqs. (46-52) on this subspace we have a Lie-algebra structure,
which is just the Lie-algebra $so(6,6, {\bf C})$.
Hence the matrices ${\Sigma}(T^{\sigma})$ for $\sigma= (001),(010),(011)$ and the matrices of the SLOCC operators
manipulating on all of the qubits except Alice's, form a representation of $so(6,6,{\bf C})$.
From the explicit form of these matrices (see Eqs. (57-58), (60-61) and (66-67)) we can see that this representation is reducible. It contains a $32$ and a $24$ dimensional representation.
The $32$ dimensional representation is the irrep of $so(6,6)$
which should be present\cite{Ferrara} in the restriction of the $56$ of $e_7$ to the maximal subalgebra $sl(2)\oplus so(6,6)$.
The $24$ dimensional representation is acting on the space
${\cal H}_{001}\oplus {\cal H}_{010}\oplus{\cal H}_{011}$ i.e. in the notation of Duff and Ferrara\cite{Ferrara} on the set $(a_{ABC}, d_{ADE}, f_{AFG})$
(see the correspondence as given by Eq.(41)).
Restriction from $E_7({\bf C})$ to the case of $E_7({\bf Z})$ this is the subsector corresponding to black hole solutions in $N=4$ supergravity with no vector multiplets.
The symmetry in this case is $SL(2, {\bf Z})\times SO(6,6,{\bf Z})$, and the black hole charges are belonging to the $(2,12)$ representation with $12$ electric and $12$ magnetic charges.
We can arrange the electric and magnetic charges in the $12$ component vectors
$p=(\psi^{001}_{0i_2i_3},\psi^{010}_{0i_4i_5},\psi^{011}_{0i_6i_7})$
and $q=(\psi^{001}_{1i_2i_3},\psi^{010}_{1i_4i_5},\psi^{011}_{1i_6i_7})$.
In this way we have obtained an explicit construction of the decomposition

\begin {eqnarray}
E_7&{\supset}& SL(2)\times SO(6,6)\nonumber\\
56&\to&(2,12)+(1,32)
\end {eqnarray}
\noindent
using the formalism of tripartite protocols of quantum information theory.

It is also clear that by looking back at the explicit form of the operators
$\Sigma(T^{001})$ as given by Eqs. (57-58) and the decomposition of Eq. (42) we can understand the further decomposition

\begin {eqnarray}
SL(2)\times SO(6,6)&\supset& SL(2)\times SL(2)\times SL(2)\times SO(4,4)         \nonumber\\
(2,12)+(1,32)&\to&(2,2,2,1)+(2,1,1,8_v)+(1,2,1,8_s)+(1,1,2,8_c).
\end {eqnarray}
\noindent
Indeed, $(2,2,2,1)$ is acting on ${\cal H}_{001}=V_1\otimes V_2 \otimes V_3$,
$(2,1,1,8_v)$ is acting on the space $V_1\otimes(V_{45}\oplus V_{67})$ with
our $8_v$ representation ${\cal R}_0$ as given by Eqs. (16) and (21) is acting on $V_{45}\oplus V_{67}$. Similarly ${\cal R}_1$ and ${\cal R}_2$ correspond to the spinor representations $8_s$ and $8_c$.
Hence $1-2-3$ (i.e. $ABC$) triality is linked with the $8_v-8_s-8_c$ triality of $SO(4,4)$ as was observed by Duff and Ferrara\cite{Ferrara}.

The explicit form of the remaining generators $\Sigma(T^\sigma)$ with $\sigma=
(100), (101), (110), (111)$ can be constructed using decompositions similar to the ones of Eq. (42), (59) and (65).
The generators not belonging to the SLOCC subalgebra are again of the (63) form where in the diagonal operators
again we have the terms $I\times {\cal D}_I$ and $I\otimes\tilde{\cal D}_I$ , $I=0,1,2$ subjected to suitable permutations which are in turn
dictated by the special form of the decomposition.
The resulting operators apart from a zero  occurring in the diagonal  corresponding to the tripartite subspace ${\cal H}_{\sigma}$ the operator  $\Sigma(T^{\sigma})$ is annihilating,  are always of the (64) form.
The order of the six terms of this kind is again fixed by the decomposition.
However, there are two important subtleties.
The first is the appearance of permutations permuting also the qubits within the superselection sectors. Hence unlike in Eqs. (58), (61) and (67) where in ${\Pi}^{\sigma}$ only the operator ${\bf 1}=I\otimes I\otimes I$ appeared, in these new cases we have also operators like $P_{123}$ permuting the qubits cyclically.
For example one can check that ${\Pi}^{100}$ is of this form where apart from the usual occurrence of ${\bf 1}$ also $P_{123}$ is appearing in the $15$ and $P_{123}^{-1}$ in the $51$ entry.
The second subtlety is to determine the particular sign the corresponding term
in the diagonal of ${\Delta}(T^{\sigma})$ has to be included. 
Looking at the explicit forms of Eq. (57), (60) and (66) we see that terms containing a $\tilde{\cal D}_I$ usually come with a negative sign. However, in Eq. (60) one of the signs had to be changed in order to get the correct commutation relation
Eq. (68). Knowing the explicit form of the  $e_7$ commutation relations (Eq.(52)) 
we can determine these signs case by case. 
Obviously the presence of these signs should be somehow related to the presence of ${\phi}(\sigma,\tau)$ appearing in Eq. (55).
We are planning to clarify such issues in a furthcoming publication.

\section{Cartan'S quartic invariant as a measure of entanglement}

Let us now consider the problem of finding an appropriate measure of entanglement for the tripartite entanglement of our seven qubits. Looking at Fig.1.
we see that there are seven tripartite systems associated to the seven lines of the Fano plane.
It is well-known\cite{Kundu}  that the unique $SL(2, {\bf C})^{\otimes 3}$ and triality invariant
measure for a three-qubit system 

\beq
\vert \psi\rangle=\sum_{A,B,C=0,1}{\psi}_{ABC}\vert ABC\rangle
\eeq
\noindent
is the three-tangle ${\tau}_3$ which is
given in terms of Cayley's hyperdeterminant by the formula

\beq
{\tau}_3\equiv 4\vert D({\psi})\vert
\eeq
\noindent
where  
\begin{eqnarray}                                                                 D({\psi})&\equiv &  {\psi}_{000}^2{\psi}_{111}^2+{\psi}_{001}^2{\psi}_{110}^2+{\psi}_{010}^2{\psi}_{101}^2+{\psi}_{011}^2{\psi}_{100}^2\nonumber\\                  &-&2({\psi}_{000}{\psi}_{001}{\psi}_{110}{\psi}_{111}+{\psi}_{000}{\psi}_{010}{\psi}_{101}{\psi}_{111}\nonumber\\&+&                                            {\psi}_{000}{\psi}_{011}{\psi}_{100}{\psi}_{111}                                +{\psi}_{001}{\psi}_{010}{\psi}_{101}{\psi}_{110}\nonumber\\&+&                  {\psi}_{001}{\psi}_{011}{\psi}_{110}{\psi}_{100}+{\psi}_{010}{\psi}_{011}{\psi}_{101}{\psi}_{100})\nonumber\\ &4&({\psi}_{000}{\psi}_{011}{\psi}_{101}{\psi}_{110}+{\psi}_{001}{\psi}_{010}{\psi}_{100}{\psi}_{111})                            \label{hypdet}                                                                  \end{eqnarray}                                                                 \noindent
or alternatively
\beq
D({\psi})=-\frac{1}{2}{\varepsilon}^{AB}{\varepsilon}^{A^{\prime}B^{\prime}}
{\varepsilon}^{CD}{\varepsilon}^{C^{\prime}D^{\prime}}
{\varepsilon}^{A^{\prime\prime}D^{\prime\prime}}{\varepsilon}^{B^{\prime\prime} C^{\prime\prime}}{\psi}_{AA^{\prime}A^{\prime\prime}}
{\psi}_{BB^{\prime}B^{\prime\prime}}{\psi}_{CC^{\prime}C^{\prime\prime}}{\psi}_{DD^{\prime}D^{\prime\prime}}.
\eeq
\noindent
Since we have seven tripartite systems we are searching for an $E_7({\bf C})$ invariant which is quartic in the amplitudes and when it is restricted to any of the subsystems corresponding to the lines of the Fano plane gives rise to Cayley's hyperdeterminant. Due to a result of Manivel (see Proposition 2. of Ref. 10.)
there is an invariant quartic form on ${\cal H}$ of Eq. (40), which is also the unique $W(E_7)$
(the Weyl group of $E(7)$) invariant quartic form, whose restriction to each tripartite system is proportional to Cayley's hyperdeterminant.
From this result it follows that this quartic invariant we are searching for should contain the sum of  {\it seven} copies of the expression of Eq\ (\ref{hypdet}), i.e. we should have the sum,
$\sum_{\sigma\neq(000)} D({{\psi}^{\sigma}})$ occurring in the explicit form of this invariant, where ${\psi}^{\sigma}\in {\cal H}_{\sigma},\quad \sigma\in {\bf Z}_2^3-(000)$.

The invariant in question is Cartan's quartic invariant $J_4$ well-known from studies concerning $SO(8)$ supergravity\cite{Cartan,Cremmer,Kol,Gunaydin}.
$J_4$ is the singlet in the tensor product representation $56\times 56\times 56\times 56$.
Its explicit form in connection with stringy black holes with their $E_{7(7)}$ symmetric area form\cite{Kol} is given either in the Cremmer-Julia form\cite{Cremmer} in terms of the complex $8\times 8$
central charge matrix $Z$ or in the Cartan form\cite{Cartan} in terms of two real $8\times 8$ ones ${\cal P}$ and ${\cal Q}$ containing the quantized electric and magnetic charges of the black hole. Its new form in terms of the $56$ {\it complex} amplitudes of our  seven qubits has been calculated in Ref. 8.
For reasons to be explained later we take the form slightly different from Ref. 8.

\begin {eqnarray}
J_4&=& \frac{1}{2}(a^4+b^4+c^4+d^4+e^4+f^4+g^4)+\nonumber\\
(a^2b^2&+&b^2c^2+c^2d^2+d^2e^2+e^2f^2+f^2g^2+g^2a^2+\nonumber\\
a^2c^2&+&b^2d^2+c^2e^2+d^2f^2+e^2g^2+f^2a^2+g^2b^2+\nonumber\\
a^2d^2&+&b^2e^2+c^2f^2+d^2g^2+e^2a^2+f^2b^2+g^2c^2)\nonumber\\
+4[bceg&+&cdef+bdfg+abef+acfg+adeg+abcd].
\end {eqnarray}
\noindent

\begin{figure}
\centerline{\resizebox{9.0cm}{!}{\includegraphics{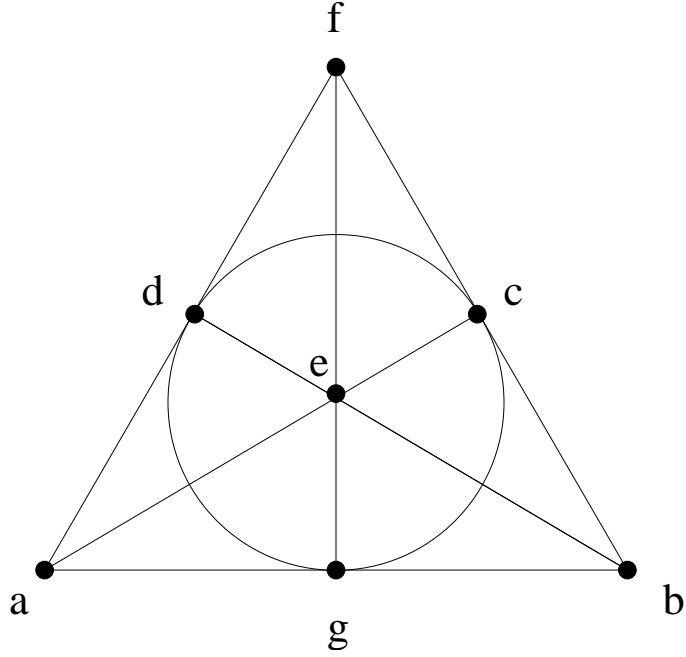}}}
\caption{\label{figr5}
The dual Fano plane. To its points now we attached tripartite states from the spaces ${\cal H}_{\sigma},\quad \sigma\in{\bf Z_2}^3-(000)$  with their complex amplitudes labelled as in Ref. 8.
See also Eq. (41). 
 } \end{figure}

Here we have for example 
\beq
a^4=-2Det(a)={\varepsilon}^{A_1A_2}{\varepsilon}^{B_1B_2}{\varepsilon}^{A_3A_4}
{\varepsilon}^{B_3B_4}{\varepsilon}^{C_1C_4}{\varepsilon}^{C_2C_3}
a_{A_1B_1C_1}a_{A_2B_2C_2}a_{A_3B_3C_3}a_{a_4B_4C_4},
\eeq
\noindent
\beq
a^2d^2={\varepsilon}^{B_1C_1}{\varepsilon}^{B_2C_2}{\varepsilon}^{D_3E_3}{\varepsilon}^{D_4E_4}{\varepsilon}^{A_1A_3}{\varepsilon}^{A_2A_4}a_{A_1B_1C_1}
a_{A_2B_2C_2}d_{A_3D_3E_3}d_{A_4D_4E_4},
\eeq
\noindent
\beq
abcd={\varepsilon}^{A_1A_4}{\varepsilon}^{B_1B_3}{\varepsilon}^{C_1C_2}{\varepsilon}^{D_3D_4}{\varepsilon}^{E_2E_4}{\varepsilon}^{F_2F_3}a_{A_1B_1C_1}b_{C_2E_2F_2}c_{B_3D_3F_3}d_{A_4D_4E_4}.
\eeq
\noindent
Notice that according to our labelling convention as given by Eq. (30) and (41)
the terms containing {\it four} tripartite systems can be written symbolically
as

\begin {eqnarray}
bceg&=&{\psi}^{101}{\psi}^{110}{\psi}^{111}{\psi}^{100}\nonumber\\
cdef&=&{\psi}^{110}{\psi}^{010}{\psi}^{111}{\psi}^{011}\nonumber\\
bdfg&=&{\psi}^{101}{\psi}^{010}{\psi}^{011}{\psi}^{100}\nonumber\\
abef&=&{\psi}^{001}{\psi}^{101}{\psi}^{111}{\psi}^{011}\\
acfg&=&{\psi}^{001}{\psi}^{110}{\psi}^{011}{\psi}^{100}\nonumber\\
adeg&=&{\psi}^{001}{\psi}^{010}{\psi}^{111}{\psi}^{100}\nonumber\\
abcd&=&{\psi}^{001}{\psi}^{101}{\psi}^{110}{\psi}^{010}\nonumber
\end {eqnarray}
\noindent 
Notice that the sum of the ${\bf Z}_2^3$ labels always gives $(000)$
corresponding to the fact that the resulting combination has no ${\bf Z}_2^3$
charge i.e. it is belonging to the singlet of $E_7$ as it has to be.
The remaining terms  of $J_4$ containing {\it two} and {\it one} tripartite states
are obviously sharing the same property.
Do not confuse however, the upper indices e.g. in ${\psi}^{001}$ 
with the lower ones occurring in Eq. (74) , e.g. ${\psi}_{001}$.
Upper indices label the superselection sectors i.e. the different types of tripartite systems and lower indices label the basis vectors of the qubits belonging to the particular triparite system.
However, it is interesting to realize that the sum of the lower indices ( regarded as elements of ${\bf Z}_2^3$) occurring in the terms of the expression for Cayley's hyperdeterminant Eq. (74) gives again $(000)$. Moreover, some of the combinations in Eq. (80) are having the same form as the ones
in Eq. (74). Indeed, the terms $abcd$, $acfg$, $bdfg$ and $adeg$
are of this form. Moreover, the terms $a^2c^2$, $b^2d^2$ and $f^4g^2$ 
can be written symbolically as $({\psi}^{001})^2({\psi}^{110})^2$,
$({\psi}^{010})^2({\psi}^{101})^2$ and , $({\psi}^{100})^2({\psi}^{011})^2$
also occurring as the second, third and fourth terms of Eq. (74).
This coincidence might be an indication that using the $56$ amplitudes in the purely ${\bf Z}_2^3$ labelled form  ${\psi}^{{\sigma}}_{ijk}$, the quartic invariant $J_4$ can be expressed in a very compact form reflecting additional symmetry properties.

Another important observation is that the terms occurring in the (76) expression for $J_4$ can be understood using the {\it dual} Fano plane.
To see this note, that the Fano plane is a projective plane hence we can use projective duality
to exchange the role of lines and planes.
Originally we attached qubits to the points, and tripartite sysems to the lines
of the Fano plane (see Figs 2. and 3.).
Now we take the dual perspective, and attach the tripartite states to the points and qubits to the lines  of the dual Fano plane (Fig. 5.).
In the ordinary Fano plane the fact that three lines are intersecting in a unique point
corresponded to the fact that any three entangled 
tripartite systems share a unique qubit.
In the dual perspective this enanglement property corresponds to the geometric one that three points are always lying on a unique line.
For example let us consider the three points corresponding to the tripartite states belonging to the subspaces ${\cal H}_{\sigma}$, 
with $\sigma =(001), (010), (011)$. According to Eq. (41) to these subspaces correspond the amplitudes $a$, $d$, and $f$.
Looking at Fig. 5. these amplitudes define the corresponding points 
lying on the line $adf$. This line is defined by the common qubit these tripartite states share i.e. qubit $A$.

In the dual Fano plane we have seven points, with  seven tripartite states attached to them. The corresponding entanglement measures are proportional to seven copies of Cayley's hyperdeterminant, then in $J_4$ we have the terms  $a^4, b^4, c^4, d^4, e^4, f^4$ and $g^4$. 
We also have seven lines with three tripartite states on each of them.
We can group the $3\times 7=21$ terms of the form $a^2b^2$ etc into seven groups associated to such lines.
They are describing the pairwise entanglement between the three different
tripartite systems (sharing a common qubit).
For example for the line $adf$ we have the terms $a^2d^2$, $a^2f^2$ and $d^2f^2$ describing such pairwise entanglements.
Finally we have seven quadrangles (as complements to the lines) with four entangled tripartite systems. They are precisely the ones as listed in Eq. (80)
giving rise to the last seven terms in $J_4$. 
Hence the terms in $J_4$ are of three type

\begin {eqnarray}
{\rm POINT}&\leftrightarrow& {\rm 1 \quad TRIPARTITE \quad STATE}               \leftrightarrow {a^4,\dots}\nonumber \\
{\rm LINE}&\leftrightarrow& {\rm 3 \quad TRIPARTITE \quad STATES}\leftrightarrow {(a^2d^2, a^2f^2, d^2f^2),\dots}\\
{\rm QUADRANGLE}&\leftrightarrow& {\rm 4 \quad TRIPARTITE\quad STATES}             \leftrightarrow {bceg,\dots}.
\end {eqnarray}
\noindent
It is useful to remember that the tripartite states forming lines are {\it sharing} a qubit, and the tripartite ones forming the  quadrangles which are complements to this line are {\it excluding} this qubit. 
For example the line $adf$ includes qubit $A$ and the complementary quadrangle
$bceg$ excludes it.

\subsection{Truncation to a line} 
We have already seen that the truncation of our system with seven tripartite states to a single tripartite one yields the three-tangle ${\tau}_3=4\vert D(\psi)\vert$ as the natural measure of entanglement. Here $\psi$ can denote any of the amplitudes from the set  ${\psi}^{\sigma}, \sigma\in {\bf Z}_2^3-(000)$.
.
In the black hole analogy where instead of the complex amplitudes of $\psi$ we  use unnormalized integer ones corresponding to the quantized  charges, the scenario we get is the one of the STU model which have already been discussed within the framework of quantum information theory\cite{Duff,Linde,Levay,Ferrara}. 
In this case the black hole entropy is given by the fomula

\beq
S=\pi\sqrt{ \vert D(\psi)\vert }=\frac{\pi}{2}\sqrt{{\tau}_3^{(1)}} 
\eeq
\noindent
i.e. it is related to the three-tangle ${\tau}_3^{(1)}$, where the upper index indicates that we have merely one tripartite system.
The geometric picture suggested by our use of the dual Fano plane is
that of a truncation of the entangled lattice to a single point.

Consider now
a truncation of the seven qubit system to one of the lines of the dual Fano plane Fig. 5.
Let us take for example the line $adf$. 
As the measure of entanglement for this case we define

\beq
{\tau}_3^{(3)}=2\vert a^4+d^4+f^4+2(a^2d^2+a^2f^2+d^2f^2)\vert,
\eeq
\noindent
where the notation ${\tau}_3^{(3)}$ indicates that now we have three tripartite states.
Notice that this invariant is proportional to the one of Duff and Ferrara\cite{Ferrara} (however, in Eq. (7.3) of Ref.4. the factor $6$ should be replaced by a factor $2$).

Now we write the state corresponding to the line $adf$ in the form
\beq
\vert\psi\rangle =\sum_{ABCDEFG=0,1}\vert A\rangle\otimes (a_{ABC}\vert BC\rangle +d_{ADE}\vert DE\rangle+f_{AFG}\vert FG\rangle).
\eeq
\noindent
This notation clearly displays that this state is an entangled one of qubit $A$
with the remaining ones $(BC)(DE)(FG)$.
Recalling that on this state the $(2,12)$ of $SL(2)\times SO(6,6)$ acts
we can write this as

\beq
\vert\psi\rangle=\sum_{A\mu}{\psi}_{A\mu}\vert A\rangle\otimes\vert \mu\rangle,\quad A=0,1,\quad \mu=1,2,\dots 12.
\eeq
\noindent
Let us discuss the role the group $SL(2)\times SO(6,6)$ plays in the quantum information theoretic context. $SL(2)$ corresponds to the usual SLOCC protocols. The second one $SO(6,6)$ contains two different types of transformations. One set corresponds to the remaining part of the SLOCC group i.e. $SL(2)^{\otimes 6}$ ($18$ generators). The other set defines transformations transforming states between the different superselection sectors. According to Section II. these transformations are generated by three sets of four-qubit states ($3\times 16$ generators). 

Denote by ${\psi}$ the $2\times 12$ matrix of Eq. (86). 
For its components ${\psi}_{A\mu}$
we introduce the notation

\beq
p^{\mu}\equiv{\psi}_{0\mu}=
\begin {pmatrix}a_{0BC}\\d_{0DE}\\f_{0FG}\end {pmatrix},\quad
q^{\mu}\equiv{\psi}_{1\mu}=
\begin {pmatrix}a_{1BC}\\d_{1DE}\\f_{1FG}\end {pmatrix}
.
\eeq
\noindent
For two $2\times 2$ complex matrices $X$ and $Y$  with components               $(X_{00},X_{01},X_{10},X_{11})$ and $(Y_{00},Y_{01},Y_{10},Y_{11})$                         let us introduce the notation                                                                                                                   \beq                                                                            X\cdot Y\equiv{\varepsilon}^{AA^{\prime}}{\varepsilon}^{BB^{\prime}}            X_{AB}Y_{A^{\prime}B^{\prime}}.                                                 \eeq                                                                            \noindent
Then for example $a^4$ and $a^2d^2$ of Eqs. (77-78) can be written in the form

\begin {eqnarray}
a^4&=&2[(a_0\cdot a_0)(a_1\cdot a_1)-(a_0\cdot a_1)^2],\nonumber \\
a^2d^2&=&(a_0\cdot a_0)(d_1\cdot d_1)+(a_1\cdot a_1)(d_0\cdot d_0)-2(a_0\cdot a_1)(d_0\cdot d_1).
\end {eqnarray}
\noindent
With the new notation

\beq
{\bf pq}\equiv h_{\mu\nu}p^{\mu}q^{\nu}=p^{\mu}q_{\mu}\equiv a_0\cdot a_1+d_0\cdot d_1+f_0\cdot f_1,\quad \mu,\nu=1,2,\dots 12
\eeq
\noindent
where the $12\times 12$ matrix $h$ with $4\times 4$ blocks as elements has the form
\beq
h=\begin {pmatrix}{\varepsilon}\otimes {\varepsilon}&0&0\\0&{\varepsilon}\otimes {\varepsilon}&0\\0&0&{\varepsilon}\otimes {\varepsilon}\end {pmatrix},
\eeq
\noindent
and the Pl\"ucker coordinates

\beq
P^{\mu\nu}\equiv p^{\mu}q^{\nu}-p^{\nu}q^{\mu},
\eeq
\noindent
we get for the invariant ${\tau}_3^{(3)}$ the following expression

\beq
{\tau}_3^{(3)}=2\vert P_{\mu\nu}P^{\mu\nu}\vert=4\vert({\bf pp})({\bf qq})-({\bf pq})^2\vert
\eeq
\noindent

In the black hole analogy using for $p^{\mu}$ and $q^{\mu}$ instead of complex numbers integers corresponding to quantized charges of electric and magnetic type the measure of entanglement in  Eq. (93) can be related to the black hole      entropy

\beq
S=\frac{\pi}{2}\sqrt{\tau_3^{(3)}},
\eeq
\noindent
coming from the truncation of the $N=8$ case with $E_{7(7)}$ symmetry to
the $N=4$ one\cite{Kallosh,Cvetic,Ferrara} with $Sl(2)\times SO(6,6)$.

From the $2\times 12$ matrix ${\psi}$ of Eq. (86) we can form the one $\varrho\equiv {\psi}{\psi}^{\dagger}$ which is just the reduced density matrix of qubit $A$ the one all of our tripartite systems share.
It is well-known\cite{Kundu,Levay2} that for normalized states $\langle\psi\vert\psi\rangle =1$ the
measure

\beq
0\leq{\tau}_{1(234567)}=4\overline{ P_{\alpha\beta}}P^{\alpha\beta}=4\vert {\rm Det}{\varrho}\vert\leq 1,\quad \alpha,\beta =1,2,\dots 12
\eeq
\noindent
gives information on the degree of separability of qubit $A$ from the rest of the system.
Here unlike in Eq. (93) summation is understood with respect to the $12\times 12$ unit matrix.
Using Eq. (95) one can prove that for normalized states

\beq
0\leq {\tau}_3^{(3)}\leq 1.
\eeq
\noindent
Indeed after noticing that

\beq
{\varepsilon}\otimes {\varepsilon}=UU^{T},\qquad U=\frac{1}{\sqrt{2}}
\begin {pmatrix}1&0&0&1\\0&i&i&0\\0&-1&1&0\\i&0&0&-i\end {pmatrix}\in SU(4)
\eeq
\noindent
with the help of $U$ we can transform the four components of the amplitudes $a_{0BC},\dots ,f_{1FG}$ to the so-called magic base\cite{Wootters} (i.e. to the base consisting of the four famous Bell-states with suitable phase factors included).
Then we have ${\tau}_3^{(3)}=2\vert P_{\alpha\beta}P^{\alpha\beta}\vert$ where  $P^{\alpha\beta}$ refers to the components of the Pl\"ucker matrix in the magic base and summation is now with respect to ${\delta}_{\alpha\beta}$. Since  $\varrho={\psi}{\psi}^{\dagger}$ is invariant with respect to this transformation $\psi\to {\psi}{\cal U}$ where ${\cal U}=U\oplus U\oplus U\in SU(6)$ the expression in Eq. (95) is not changed. 
Using Eq. (95) and the triangle inequality 

\beq
0\leq 4\vert P_{\alpha\beta}P^{\alpha\beta}\vert\leq 4\overline{P_{\alpha\beta}}P^{\alpha\beta}\leq 1,
\eeq
\noindent
hence we get Eq. (96).
An immediate consequence of this is that
${\tau}_3^{(3)}$ vanishes for systems where qubit $A$ is separable from the rest. Similar conclusions can be drawn from the vanishing of the ${\it six}$ quantities (based on the remaining six qubits) defined accordingly.
The six new quantities ${\tau}_3^{(3)}$  are vanishing when any qubit located at the vertices of the Fano plane is separable from the tripartite systems
associated with the three lines the qubit is lying on.
It is important to realize however, that one can also get ${\tau}_3^{(3)}=0$ 
by choosing $a_{100}=a_{010}=a_{001}=d_{100}=d_{010}=d_{001}=f_{100}=f_{010}=f_{001}=1/3$. This state corresponds to the situation of chosing three different tripartite states belonging to the Werner class. 
These tripartite states are genuine entangled three-qubit ones which retain maximal bipartite entanglement when any one of the three qubits is traced out\cite{Vidal}.

\subsection{Truncation to a quadrangle}

Having discussed the truncation to a line of the dual Fano plane,               now we consider 
the complementary situation, i.e. truncation to a quadrangle.
By a quadrangle as usual we mean the complement of a line.
We have seen that there is a complementary relationship between the entanglement properties as well.
Three tripartite systems associated to a line share a common qubit, and four tripartite systems associated to the complement of this line exclude precisely this qubit. Hence we are expecting this relationship to be manifest in the special form of an entanglement measure characterizing this situation.

As an example let us consider again the line $adf$ and its complement the quadrangle $bceg$.  We define the quantity

\beq
{\tau}_3^{(4)}=2\vert b^4+c^4+e^4+g^4+2(b^2c^2+b^2e^2+b^2g^2+c^2e^2+c^2g^2+e^2g^2)+8bceg\vert.
\eeq
\noindent
Here the notation ${\tau}_3^{(4)}$ refers to the situation of entangling {\it four} tripartite systems.
In the following we prove that ${\tau}_4^{(4)}$ is the entanglement measure characterizing
the configuration complementary to the one of the previous subsection.

First let us recall few facts supporting our claim.
The first observation is a group theoretic one. The amplitudes $a$, $d$ and $f$
distributed in the form of Eq. (87) transform according to the (2,12) of $SL(2)\times SO(6,6)$. The complementary amplitudes $b$, $c$, $e$ and $g$ are transforming according to the $(1,32)$ of $SL(2)\times SO(6,6,)$ 
, i.e. they are spinors under $SO(6,6)$. 
This fact is clearly displayed in our explicit matrix representation Eqs. (57-58), (60-61) and (66-67). (See also the correspondence of Eq. (41).)
Hence our invariant ${\tau}_3^{(4)}$ should also be regarded as the singlet in the symmetric tensor product of $4$ spinor representations of $SO(6,6)$.

Our second observation is 
based on the black hole analogy.
Let us relate our (unnormalized) amplitudes $a,b,\dots g$ to the quantized charges of the $E_{7(7)}$ symmetric area form\cite{Kol} of the black hole. In this case we have $7\times 8=56$ integers regarded as amplitudes of a seven qubit system associated to the entangled lattice defined by the Fano plane. These amplitudes correspond to the two $8\times 8$ antisymmetric
matrices of charges ${\cal P}$ and ${\cal Q}$.
Then the Cartan form of our quartic invariant $J_4({\cal P}, {\cal Q})$ is\cite{Cartan}

\beq
J_4({\cal P}, {\cal Q})=-{\rm Tr}({\cal Q}{\cal P}{\cal Q}{\cal P})+\frac{1}{4}\left({\rm Tr}{\cal Q}{\cal P}\right)^2-4\left({\rm Pf}({\cal P})+{\rm Pf}({\cal Q})\right).
\eeq
\noindent
In the context of toroidal compactifications of M-theory or  type II string theory the antisymmetric matrices ${\cal P}$ and ${\cal Q}$
may be indentified as\cite{Pioline}

\beq
{\cal Q}=\begin {pmatrix}[D2]^{mn}&[F1]^m&[kkm]^m\\-[F1]^m&0&[D6]\\-[kkm]^m&-[D6]&0\end {pmatrix},\quad
{\cal P}=\begin {pmatrix}[D4]_{mn}&[NS5]_m&[kk]_m\\-[NS5]_m&0&[D0]\\-[kk]_m&-[D0
]&0\end {pmatrix}\quad m,n=1,\dots 6.
\eeq
\noindent
Here, $[D2]^{mn}$ denotes a $D2$ brane wrapped along the directions $mn$ of a six dimensional torus $T^6$.
$[D4]_{mn}$ corresponds to $D4$-branes wrapped on all directions {\it but} $mn$, $[kk]_m$ denotes a momentum state along direction $m$, $[kkm]^m$ a Kaluza-Klein $5$-monopole localized along the direction $m$, $[F1]^m$ a fundamental string winding along direction $m$, and $[NS5]_m$ a $NS5$-brane wrapped on all directions but $m$. 

Then the $N=4$ truncation where

\beq
{\cal Q}=\begin {pmatrix}0&[F1]^m&[kkm]^m\\-[F1]^m&0&0\\-[kkm]^m&0
&0\end {pmatrix},\quad
{\cal P}=\begin {pmatrix}0&[NS5]_m&[kk]_m\\-[NS5]_m&0&0\\-[kk]_m&0
&0\end {pmatrix},
\eeq
\noindent
should corresponds to the case of our truncation to a line (e.g. the one $adf$).
In this case our ${\tau}_3^{(3)}$ is just the quartic invariant with respect to $SL(2)\times SO(6,6)$.  

The complementary case

\beq
{\cal Q}=\begin {pmatrix}[D2]^{mn}&0&0\\0&0&[D6]\\0&-[D6
]&0\end {pmatrix},\quad
{\cal P}=\begin {pmatrix}[D4]_{mn}&0&0\\0&0&[D0]\\0&-[D0
]&0\end {pmatrix},
\eeq
\noindent
of the $N=2$ truncation should correspond to our restriction to quadrangles (e.g. the one $bceg$).
The resulting quartic invariant, also based on the Jordan algebra\cite{Pioline}
$J_3^{\bf H}$ should be related to our ${\tau}_3^{(4)}$.

For an explicit proof of our                                                    claim what we need is a precise correspondence 
between the amplitudes $a,b,\dots g$ and the components of ${\cal P}$ and ${\cal Q}$.
This would also establish an explicit connection between our $56$ of $E_7$
in terms of seven qubits and the one of Cartan\cite{Cartan} in terms of the antisymmetric matrices ${\cal P}$ and ${\cal Q}$.

In order to prove our claim by establishing this correspondence we proceed as follows. We already know that our expression for the entanglement measure associated with $J_4$ should give the three-tangle Eq. (73) when restricting to a point of the dual Fano plane. Let us consider this point to be $g$ i.e. the amplitude  $g_{CDG}$ for the three-qubit state is the one of Charlie, Daisy and George.
We arrange the $2\times 4$ complex amplitudes of $g_{CDG}$ in ${\cal Q}$ and ${\cal P}$
as follows

\beq
{\cal P}=\begin {pmatrix}g_{001}&0&0&0\\0&g_{010}&0&0&\\0&0&g_{100}&0\\0&0&0&g_{111}\end {pmatrix}\otimes {\varepsilon},\qquad
{\cal Q}=\begin {pmatrix}g_{110}&0&0&0\\0&g_{101}&0&0&\\0&0&g_{011}&0\\0&0&0&g_{000}\end {pmatrix}\otimes {\varepsilon}.
\eeq
\noindent
Then from Eq. (100) using Eq. (74) we get

\beq
J_4=-D(g)=\frac{1}{2}g^4,
\eeq
\noindent
in accordance with our formula Eq. (76) and the result of Kallosh and Linde\cite{Linde}. Hence it is natural to define a  normalized measure of entanglement for our seven qubit system as

\beq
{\tau}_7\equiv 4\vert J_4\vert.
\eeq
\noindent
Indeed, for normalized states truncation to a single tripartite system gives rise to 
the three-tangle ${\tau}_3$ satisfying the constraint $0\leq{\tau}_3\leq 1$.
Moreover, for the important special case of putting GHZ states to the seven vertices of the dual Fano plane ($a_{000}=a_{111}=b_{000}=\dots =g_{111}=1/\sqrt{14}$) we get ${\tau}_7=1$.

In the black hole analogy however, the amplitudes are integers and no normalization condition is used. The special case having only $g\neq 0$ in Eq. (104) is the case of the STU model\cite{Duff,Linde,Levay}.
Notice that the amplitudes $g_{CDG}$ are occurring as the entries in the canonical form of the antisymmetric matrices ${\cal P}$ and ${\cal Q}$. This is due to the fact that our choice of the parties Charlie, Daisy and George is special. Hence we expect that this special choice will be reflected in our
choice for filling in the missing entries
of the matrices ${\cal P}$ and ${\cal Q}$ in the general case.

For normalized states truncation to the tripartite systems $adf$ lying on a line of the dual Fano plane we choose

\beq
{\cal P}=\begin {pmatrix}0&0&0&a_0^T\\0&0&0&d_0^T\\0&0&0&f_0^T\\-a_0&-d_0&-f_0&0\end {pmatrix},\qquad
{\cal Q}=\begin {pmatrix}0&0&0&-\tilde{a}_1\\0&0&0&-\tilde{d}_1\\0&0&0&-\tilde{f}_1\\\tilde{a}_1^T&\tilde{d}_1^T&\tilde{f}_1^T&0\end {pmatrix}.
\eeq
\noindent
Here the elements of these matrices are $2\times 2$ matrices constructed as follows. As we have stressed in our chosen arrangement the role of qubits $C$, $D$, and $G$ are special.
These qubits are contained in the corresponding three-qubit amplitudes
$a_{ABC}$, $d_{ADE}$ and $f_{AFG}$.
We split the $8$ components of these amplitudes into two $2\times 2$ matrices
based on the positions of the special qubits they contain

\beq
a_0=a_{AB0},\quad a_1=a_{AB1},\quad d_0=d_{A0E},\quad d_1=d_{A1E},\quad
f_0=f_{AF0},\quad f_1=f_{AF1}.
\eeq
\noindent
Moreover the Wootters spin flip operation already used in Eq. (14) is

\beq
\tilde{M}={\sigma}_2M^T{\sigma}_2=-{\varepsilon}M^T{\varepsilon}. 
\eeq
\noindent

In order to check that $4\vert J_4({\cal P}, {\cal Q})\vert$ of Eq. (100) restricted to the line $adf$ with components as given by Eq. (107) indeed gives back
our expression for ${\tau}_3^{(3)}$ (see Eq. (84)) we note that in this case  we can write $J_4$ in the form

\beq
J_4=4{\rm Det}(X^TY)-({\rm Tr}(X^TY))^2,\qquad X=\begin {pmatrix}a_0^T\\d_0^T\\f_0^T\end {pmatrix},\quad Y=\begin {pmatrix}\tilde{ a}_1\\ \tilde{d}_1\\ \tilde{f}_1\end {pmatrix}.
\eeq
\noindent
Using the identity valid for $2\times 2$ matrices

\beq
{\rm Det}(A+B)={\rm Det}A+{\rm Det}B+{\rm Tr}(A\tilde{B})
\eeq
\noindent
and grouping the terms we
get
\begin {eqnarray}
J_4&=&[4{\rm Det}(a_0\tilde{a}_1)-({\rm Tr}(a_0\tilde{a}_1))^2] 
+[4{\rm Det}(d_0\tilde{d}_1)-({\rm Tr}(d_0\tilde{d}_1))^2]                      +[4{\rm Det}(f_0\tilde{f}_1)-({\rm Tr}(f_0\tilde{f}_1))^2]\nonumber\\
&+&4{\rm Tr}(a_0\tilde{a}_1d_1\tilde{d}_0)+
4{\rm Tr}(a_0\tilde{a}_1f_1\tilde{f}_0)+
4{\rm Tr}(d_0\tilde{d}_1f_1\tilde{f}_0)\\
&-&2{\rm Tr}(a_0\tilde{a}_1)
{\rm Tr}(d_0\tilde{d}_1)
-2{\rm Tr}(a_0\tilde{a}_1)
{\rm Tr}(f_0\tilde{f}_1)
-2{\rm Tr}(d_0\tilde{d}_1)
{\rm Tr}(f_0\tilde{f}_1).
\nonumber
\end{eqnarray}
The first three terms gives {\it minus} the Cayley hyperdeterminants
$-D(a)$, $-D(d)$ and $-D(f)$, i.e. the terms $a^4/2$, $d^4/2$ and $f^4/2$.
These terms are permutation invariant hence our singleing out of qubits $C$, $D$ and  $G$
in obtaining their special forms is not relevant.
However, in our previous derivation of terms like the ones $a^2d^2$ in Eq. (89) playing a role in the invariant ${\tau}_3^{(3)}$ we have singled out the {\it first} qubit $A$ which is common to all of our tripartite states.
But now, in the expressions as given by Eq. (112) for the last six terms qubits $C$, $D$ and $G$ are playing a special role.  
We can relate the different descriptions by using the identity

\beq
4{\rm Tr}(a_0\tilde{a}_1d_1\tilde{d}_0)-2{\rm Tr}(a_0\tilde{a}_1){\rm Tr}(d_0\tilde{d}_1)=a^2d^2
\eeq
\noindent
as can be checked by a straightforward calculation, and similar ones for the terms $a^2f^2$ and $d^2f^2$.
These considerations give the result

\beq
4\vert J_4({\cal P},{\cal Q})\vert ={\tau}_3^{(3)}
\eeq
\noindent
where ${\cal P}$ and ${\cal Q}$ is given by Eq. (107) as claimed.

Finally we consider the complementary situation i.e. restriction to the quadrangle $bceg$.  
Let us consider the matrices

\beq
{\cal P}=\begin {pmatrix}g_{001}{\varepsilon}&b_0^T&c_0^T&0\\-b_0&g_{010}{\varepsilon}&e_0^T&0\\-c_0&-e_0&g_{100}{\varepsilon}&0\\0&0&0&g_{111}{\varepsilon}\end {pmatrix},\qquad
{\cal Q}=\begin {pmatrix}g_{110}{\varepsilon}&-\tilde{b}_1&-\tilde{c}_1&0\\\tilde{b}_1^T&g_{101}{\varepsilon}
&-\tilde{e}_1&0\\\tilde{c}_1^T&\tilde{e}_1^T&g_{011}{\varepsilon}&0\\0&0&0&g_{000}{\varepsilon}\end {pmatrix}.
\eeq
\noindent
Here the elements of these matrices are again $2\times 2$ matrices.
The tripartite systems with amplitudes $b_{CEF}$, $c_{BDF}$ and $e_{BEG}$ are again containing our special qubits $C$, $D$ and $G$.
The matrices occurring in the entries of ${\cal P}$ and ${\cal Q}$ are

\beq
b_0=b_{0EF},\quad b_1=b_{1EF},\quad c_0=c_{B0F},\quad c_1=c_{B1F},\quad
e_0=e_{BE0},\quad e_1=e_{BE1}.
\eeq
\noindent

Then a straightforward but tedious calculation shows that using ${\cal P}$ and ${\cal Q}$ of Eq. (115) we get

\beq
4\vert J_4({\cal P}, {\cal Q})\vert ={\tau}_3^{(4)},
\eeq
\noindent
where ${\tau}_3^{(4)}$ is given by the expression of Eq. (99).
In our derivation we have used the formula

\begin {eqnarray}
{\rm Det}{\cal P}=({\rm Pf}{\cal P})^2&=&(g_{100}g_{010}g_{001}g_{111}-g_{111}g_{100}{\rm Det}(b_0)-g_{111}g_{010}{\rm Det}(c_0)\nonumber\\&-&g_{111}g_{001}{\rm Det}(e_0)+
g_{111}{\varepsilon}^{BB^{\prime}}{\varepsilon}^{EE^{\prime}}{\varepsilon}^{FF^{\prime}}b_{0EF}c_{F^{\prime}0B^{\prime}}e_{BE^{\prime}0})^2.
\end {eqnarray}
\noindent

Using these results it is clear now that in the black hole analogy truncation to a line of our entangled system corresponds to the one of truncating the $N=8$
case with moduli space $E_{7(7)}/SU(8)$ to the $N=4$ one with moduli space $(SL(2)/U(1))\times (SO(6,6)/SO(6)\times SO(6))$.
Moreover, the truncation to a quadrangle complementary to this line gives rise
to the $N=2$ truncation\cite{Sierra} with the moduli space being $SO^{\ast}(12)/U(6)$.
It is also known\cite{Belucci} that the manifold $SO^{\ast}(12)/U(6)$ is the largest one
which can be obtained as a consistent truncation of the $N=8$, $d=4$ supergravity based on $E_{7(7)}/SU(8)$.

\section{Conclusions}

In this paper we have studied an entangled quantum system consisting of tripartite subsystems built from seven qubits. 
We have shown that using this system it is possible to extend the multiple relations found between quantum information theory and the physics of four dimensional $N=2$ stringy black holes\cite{Duff,Linde,Levay,Ferrara} to the more general $N=8$ case with $E_{7(7)}$ symmetry.

As a first step we investigated
the properties of this unusual type of entanglement and realized that they are  encoded into the discrete geometry of the Fano plane the smallest projective plane.
This entanglement is in turn intimately connected to the geometry of the group $E_7({\bf C})$.
In the arising geometric picture the basic objects of the Fano plane, namely points, lines and quadrangles (complements to lines) associated to qubits, three and four qubit systems
respectively, are forming an entangled lattice.
To the seven tripartite systems corresponding to the lines we can assign a ${\bf Z }_2^3$ charge characterizing different superselection sectors.
It has turned out that the four-qubit states associated to the quadrangles
describe $112$ from the infinitesimal operators of the Lie algebra $e_7({\bf C})$. They generate transformations connecting the different superselection sectors. Alternatively, they correspond to quantum information theoretic protocols
relating the different tripartite subsystems. The remaining $21$ ones form the infinitesimal operators of an $sl(2, {\bf C})^{\oplus 7}$ algebra. 
They generate stochastic local operations and classical communication (SLOCC)
transformations {\it within} the tripartite subsystems. 
We have built the $56$ dimensional complex vector space corresponding to the fundamental representation of $E_7({\bf C})$ as a direct sum of the Hilbert spaces of these tripartite subsystems.
We have sketched a method for the construction of the action of the $e_7({\bf C})$ infinitesimal operators (regarded as generators of tripartite protocols) 
on this space.
We have found that the details of this action are also encoded into the geometry of the Fano plane. The code which encapsulates the properties of the entangled system  
and the structure of this representation associated to it,  is just the Hamming code known from studies concerning error correction codes in quantum information theory. 

As a next step following the insight of Duff and Ferrara\cite{Ferrara} we have shown that  a natural measure of entanglement ${\tau}_7$ characterizing our entangled lattice  is provided by the quartic Cartan-Cremmer-Julia invariant $J_4$.
By using the {\it dual} Fano plane we illuminated the physical meaning of the terms occurring in this invariant.
These terms are describing the entanglement properties of three different types of subsystems namely ones consisting of one, three or four tripartite systems.
For the quantification of different types of partial entangement we         introduced the corresponding measures ${\tau}_3^{(1)}$ , ${\tau}_3^{(3)}$ and 
${\tau}_3^{(4)}$.

These results enabled us to extend further the so called black hole analogy\cite{Duff,Linde,Levay,Ferrara}.
In this analogy we consider special types of {\it unnormalized} entangled states with integer-valued amplitudes coming from the quantized electric and magnetic charges occurring in black hole solutions found in four dimensional string theory. 
Then one finds that for such systems the natural measures of entanglement can
be related to the macroscopic entropy of black holes expressed in terms of      the conserved charges.
Moreover, the classification of black hole solutions of more general type
(BPS and non-BPS)
are related to the classification of entangled states\cite{Linde}.
This analogy has already provided insights in the case of $N=2$ supersymmetric black hole solutions in the STU model.
Here we have shown that the analogy can be extended to the more general case
of black hole solutions in $N=8$ supergravity.
The scalar manifold (moduli space) in this case is $E_{7(7)}/SU(8)$, and the
black hole solutions give rise to an $E_{7(7)}$ symmetric entropy formula\cite{Kol}
expressed in terms of the $56$ charges.
These charges in turn can be regarded as the ones coming from toroidal compactification of $M$-theory or type II string theory.
The $56$ charges in this case can be related to $[D0]$, $[D2]$, $[D4]$, $[D6]$
and $[NS5]$ branes, momentum states, fundamental string windings, and Kaluza-Klein $5$-monopoles localized along the different directions of the six torus (see Eq. (101)).
We managed to relate these quantities to the integer valued amplitudes of our
unnormalized entangled state (see Eqs. (107) and (115)).
We have seen that different types of truncations of the $N=8$ case correspond to different types of entangled sublattices of the Fano plane.
In particular we have shown that there is a dual relationship between truncation to a line or truncation to its complement.
This duality relates subsystems consisting of three and four tripartite systems.
The former subsystems contain a common qubit the latter ones exclude this qubit.
The corresponding picture in the black hole analogy is the one of
obtaining the $N=4$ and $N=2$ truncations
 with moduli spaces $(SL(
 2)/U(1))\times (SO(6,6)/SO(6)\times SO(6))$
and $SO^{\ast}(12)/
 U(6)$ respectively.
The corresponding black hole solutions possess entropy formulas with $24$ and $32$ charges. They are having the form of Eq. (94) containing our entanglement measures ${\tau}_3^{(3)}$ or ${\tau}_3^{(4)}$.

Notice that these results uncovered a symmetry which could be similar to the one responsible to the string triality picture of Ref.25.
We have seven different ways to do any of the truncations to points, lines and quadrangles. These are different ways to get the $N=2$ (STU) truncation with $8$, the $N=4$ one with $24$,
and the $N=2$ one with $32$ charges. These seven ways also correspond to our seven different ways of filling in the entries of the matrices ${\cal P}$ and ${\cal Q}$. In Eqs. (107) and (115) we have chosen a one based on a special role attached to the tripartite system of Charlie, Daisy and George.
It would be nice to understand this symmetry clearly displayed in the entanglement picture also within the framework of $U$-duality\cite{Townsend}.

Untill this point in the black hole analogy we have used merely entangled systems characterized by unnormalized states with integer amplitudes related to the quantized charges.
However, in the STU model we have shown\cite{Levay} that it is useful to look at this model using {\it complex} entangled states which are local unitary equivalent to ones with {\it real} amplitudes. In the case of the STU model these states (rebits) are of the form 

\beq
\vert\Psi(S,T,U)\rangle={\cal A}(S)\otimes{\cal B}(T)\otimes {\cal C}(U)\vert\psi\rangle,
\eeq
\noindent
where $S$ , $T$ and $U$ are the {\it complex} moduli fields occurring in the STU-model, and $\vert\psi\rangle$ is the unnormalized three-qubit state with integer amplitudes related to the $8$ charges.
The $2\times 2$ matrices ${\cal A}(S)$, ${\cal B}(T)$ and ${\cal C}(U)$ depending on the moduli are elements of the SLOCC group $SL(2, {\bf C})^{\otimes 3}$.
These entangled rebits that are composites of the moduli and the charges
have been optimized with respect to $S$, $T$ and $U$ to obtain a quantum information theoretic version of the attractor mechanism\cite{Kallosh}.
It turned out that the state $\vert\Psi\rangle$ calculated at the {\it frozen} values of the moduli is a maximally entangled $GHZ$-state.

Can we have a similar generalization in the $N=8$ case as well?
In this case we already know that the protocols are again tripartite ones though we have to supplement the set of SLOCC transformations in Eq. (119) with the    ones of Eq. (64) operating {\it between} the different tripartite states.
Now the moduli space is the $70$ dimensional one $E_{7(7)}/SU(8)$.
In order to write down a real quantum bit version of Eq. (119) one should clarify the nature of the {\it real} states associated with the group $E_{7(7)}$
within the {\it complex ones} related to the group $E_7({\bf C})$.
Using the explicit constructions as given by this paper this can in principle be done.
Then we conjecture that the solutions\cite{Ferrara2} to the attractor equations obtained by finding the critical points of the black hole potential in the $N=8$ case can be understood as the process of maximization of entanglement for our  entangled lattice as defined by the Fano plane.

Can we generalize our analogy even further? It is quite natural to expect
that black hole solutions associated with magic supergravities are viable candidates for a possible generalization\cite{Ferrara3}. 
For this generalization to work one should have an underlying entangled system
of some number of qubits. Or, in mathematical terms one should be able to build up the Lie-algebras occurring in Freudenthal's Magic Square in terms of $SL(2, {\bf C})$ modules, and to find suitable representation spaces for them. 
Due to the results of Elduque\cite{Elduque} this might be done for the split form of such Lie-algebras.
Indeed, based on this result all the Lie algebras in Freudenthal's Magis Square can be constructed in a unified way using copies of SLOCC transformations and
of its natural module provided by qubits.

Finally we remark that the qubit picture of $E_7({\bf C})$ as developed in this paper is quite natural. 
After all its fundamental representation can be written in terms of $7\times 7$
matrices, though with components taken from the set of admissible tripartite
protocols on seven qubits!

\section{Acknowledgements}                                                      Financial support from the Orsz\'agos Tudom\'anyos Kutat\'asi Alap              (grant numbers T047035, T047041, T038191) is                                    gratefully acknowledged.

\end {document}